\documentclass[modern]{aastex631}

\usepackage{BobsPreamble}

\newcommand*{\img}[1]{%
    \raisebox{-.05\baselineskip}{%
        \includegraphics[
        height=0.65\baselineskip,
        keepaspectratio,
        ]{#1}%
    }%
}

\begin{document}

\title{Cholla-MHD: An Exascale-Capable Magnetohydrodynamic Extension to the Cholla Astrophysical Simulation Code}

\author[0000-0002-4475-3181]{Robert V. Caddy}
\affiliation{University of Pittsburgh, Department of Physics \& Astronomy \\
3941 O'Hara St \\
Pittsburgh, PA 15260, USA}

\author[0000-0001-9735-7484]{Evan E. Schneider}
\affiliation{University of Pittsburgh, Department of Physics \& Astronomy \\
3941 O'Hara St \\
Pittsburgh, PA 15260, USA}

\begin{abstract}

    We present an extension of the massively parallel, GPU native, astrophysical hydrodynamics code Cholla to magnetohydrodynamics (MHD).
    Cholla solves the ideal MHD equations in their Eulerian form on a static Cartesian mesh utilizing the Van Leer + Constrained Transport integrator, the HLLD Riemann solver, and reconstruction methods at second and third order.
    Cholla's MHD module can perform $\approx260$ million cell updates per GPU-second on an NVIDIA A100 while using the HLLD Riemann solver and second order reconstruction.
    The inherently parallel nature of GPUs combined with increased memory in new hardware allows Cholla's MHD module to perform simulations with resolutions $\sim500^3$ cells on a single high end GPU (e.g. an NVIDIA A100 with 80GB of memory).
    We employ GPU direct MPI to attain excellent weak scaling on the exascale supercomputer \textit{Frontier}, while using 74,088 GPUs and simulating a total grid size of over 7.2 trillion cells.
    A suite of test problems highlights the accuracy of Cholla's MHD module and demonstrates that zero magnetic divergence in solutions is maintained to round off error.
    We also present new testing and continuous integration tools using GoogleTest, GitHub Actions, and Jenkins that have made development more robust and accurate and ensure reliability in the future.

    \end{abstract}

\section{Introduction}
\label{sec:intro}



Over the past decade it has become increasingly clear that magnetohydrodynamics (MHD) plays a significant role in a variety of astrophysical phenomena \citep[e.g.][]{charbonneau_2014, beck_2016, naab_2017, han_2017, kaspi_2017, blandford_2019, Nakariakov_2020, davis_2020, Trujillo_2022, Philippov_2022, brandenburg_2023}. Magnetic fields couple to gas both directly through plasma interactions with the magnetic field, and indirectly through cosmic ray transport \citep[e.g.][]{pfrommer_simulating_2017, girichidis_spectrally_2019, chan_cosmic_2019, buck_effects_2020, werhahn_cosmic_2021, yoshida_trajectory_2021, girichidis_spectrally_2022, nunez-castineyra_cosmic-ray_2022, werhahn_gamma-ray_2023}, anisotropic conduction \citep[e.g.][]{yang_fermi_2012, hanasz_cosmic_2013, simpson_role_2016, girichidis_launching_2016, pakmor_galactic_2016, bruggen_2023}, and other physical effects.

Many studies have focused on the role of magnetic fields in galaxy evolution and have provided intriguing hints as to the possible effects of magnetic fields in galaxy dynamics \citep{shin_2008, banda_2016, grand_auriga_2017, pakmor_magnetic_2017, hopkins_but_2020, wibking_2021}. Different simulations employing different numerical methods show varying impacts of magnetic fields, ranging from magnetic fields being largely irrelevant on large scales to magnetic fields being critically important in determining the evolution and structure of the interstellar medium (ISM), modifying galactic feedback, and influencing the structure of the circumgalactic medium (CGM) \citep{martin-alvarez_how_2020, dobbs_magnetic_2007, kim_vertical_2015, hanasz_global_2009, pakmor_simulations_2013, banda_2017}. One factor in this uncertainty is the effect of numerical resolution on MHD simulations -- galactic magnetic fields are likely amplified by a turbulent dynamo, which operates over a large dynamic range \citep{martin-alvarez_three-phase_2018, beck_1996, gent_2021, carteret_2022, brandenburg_2023}. The accuracy with which the dynamo is captured thus depends on both the numerical method that is employed as well as the resolution. Thus, a simple way to extend the dynamic range captured in an MHD simulation is by employing very high resolution MHD simulations -- simulations that are now possible thanks to recent developments in hardware, numerical algorithms, and software.

Modern numerical methods for MHD are sophisticated and robust, but even with highly optimized codes, MHD simulations remain very computationally expensive \citep{athena++_2020}. This computational expense is a result of the high number of floating point calculations required by modern finite-volume methods and the heavy memory bandwidth demands of MHD codes \citep{k_athena_2021}. In addition, MHD turbulent dynamos operate across a large dynamic range, from the full scale of a galaxy all the way down to a few parsecs or smaller, five or more orders of magnitude spatially \citep{pariev_magnetic_1_2007, pariev_magnetic_2_2007, ntormousi_dynamo_2020, galishnikova_tearing_2022}. As a result, high resolution simulations are critical in order to accurately capture the effects of magnetic fields in astrophysical simulations.

This need has driven a push to develop MHD codes that can take advantage of modern computer architectures \citep[e.g.][]{schive_gamer-2_2018, almgren_castro_2020, zingale_castro_2020, shankar_gram-x_2022, liska_h-amr_2022, begue_cuharm_2023, holmen_early_2023, parthenon_2023}. Very large, high resolution MHD simulations require supercomputers to run due to their high computational cost. The primary source of computational power in most new supercomputers is Graphics Processing Units (GPUs)\footnote{\url{https://www.top500.org/lists/top500/2023/11/}}. For example, of the top ten supercomputers named in the November 2023 Top500 list, only one - Fugaku - does not rely on GPUs for the majority of its performance.\footnote{Fugaku employs custom CPUs that utilize vector processors similar to GPUs for much of its performance.} The ubiquity of GPUs in modern supercomputers thus necessitates the development of GPU-based astrophysical MHD simulation codes.

The induction equation implies a simple constraint; it requires that the magnetic field be divergence free, i.e. the Universe does not contain magnetic monopoles. However, achieving this constraint numerically is not trivial, and several methods have been developed. Among these are the Powell 8 Wave scheme, which adds an additional source term to the induction equation and uses an 8 wave Riemann solver with the 8th wave corresponding to the magnetic divergence \citep{Powell1997};
vector potential/projection methods, which project the magnetic field into the scalar and vector potential and then perform a cleaning step to reduce the divergence to zero by solving a Poisson equation \citep{brackbill_1980,ryu_1995,Crockett_2005,torrilhon2005locally};
divergence cleaning methods, which operate by modifying the system of conserved equations with corrections that dissipate and propagate out the magnetic divergence \citep{dedner_hyperbolic_2002,mignone_2010}; and constrained transport (CT), which evolves face-centered values of the magnetic field and updates them using the electromotive forces \citep{evans_1988}.
We have chosen to implement the CT method in Cholla due to its overall accuracy, because the algorithm pairs well with static structured grids, and because the computational efficiency of GPUs is a good match to the algebraic complexity of CT.

Constrained transport is formally divergence free, and when implemented numerically it typically results in divergence errors on the order of machine round off error \citep{evans_1988, gardiner_2005, stone_athena_2008, stone_2009, zingale_castro_2020, almgren_castro_2020}. This is accomplished by tracking magnetic fields on a staggered, face centered grid rather than using cell-centered averages. These face centered values are used in conjunction with Riemann fluxes to calculate edge centered electric fields, and those electric fields are used to update the magnetic field. Thus, the trade-off for a divergence-free method is significant additional algorithmic complexity and associated computational expense, which can make the method more challenging to implement in a particle based scheme or when using unstructured meshes.

The Cholla code (Computational Hydrodynamics On paraLLel Architectures) \citep{schneider_2015} is a fixed grid, finite volume hydrodynamics code for astrophysics that was designed to run natively on GPU-based supercomputers. It employs an unsplit 3D hydrodynamics integrator based on the Van Leer predictor-corrector method \citep{falle_1991, van_leer_2006} and was designed to be extended to MHD using constrained transport \citep{evans_1988}. This work presents the MHD extension of Cholla. Our MHD implementation largely follows the Van Leer + Constrained Transport (VL+CT) method presented in \cite{stone_2009} with modifications for GPUs. It also uses an HLLD Riemann solver \citep{hlld_2005} and includes second \citep{stone_2009} and third \citep{felker_2018} order reconstruction in the characteristic variables \citep{stone_athena_2008}. We also highlight implementation choices that are particularly relevant to solving these equations on GPUs.

The extension of Cholla to include MHD allows the simulation of previously unreachable domains. The VL+CT integrator provides high accuracy results with divergences that are zero to round off error. Given current memory constraints, the code is fast enough that a $\sim459^3$ cell MHD simulation can be run on a single AMD MI250X Graphics Compute Die (GCD) (logically a single GPU), allowing high resolution simulations to be run with only a small number of local resources. In addition, Cholla scales up to power of the largest available supercomputers, enabling simulations up to $19,278^3 \approx 7.2$ trillion cells on \textit{Frontier}\footnote{\url{https://www.olcf.ornl.gov/frontier/}}, the world's first exascale supercomputer. This will allow MHD simulations of entire galaxies with a constant resolution of a few parsecs per cell, turbulent box simulations with many trillions of cells, or many thousands of lower resolution simulations to be computed rapidly. For example, with approximately the same computing resources used to evolve a $19,278^3$ simulation one could run thousands of $2,000^3$-cell simulations, enabling entire parameter studies with resolutions comparable to current cutting edge CPU-based simulations.

In addition to requiring complex algorithms to produce accurate results, modern community-developed simulation codes like Cholla require robust testing and software-development infrastructure to maintain their reliability. This work also presents the implementation of an automated testing/continuous integration (CI) pipeline for Cholla. CI tools have expanded rapidly in functionality and popularity over the last 20 years and their usefulness in scientific software is well established \citep{beck_1999, wilson_2014,wilson_2017}. Particularly in the last 5 years with the advent of GitHub Actions, Jenkins, and similar easily accessible and cheap (or even free) tools, CI pipelines have become much more straightforward to set up and run even for small groups and individuals. We present our implementation of testing and CI that is designed to be straightforward and scalable from a single GPU all the way up to an exascale machine.

The outline of this paper is as follows. In Section \ref{sec:methods}, we describe our implementation of the VL+CT algorithm in detail along with the modifications we made to efficiently run on GPUs. In Section \ref{sec:mhd-tests}, we demonstrate the correctness and accuracy of Cholla on a suite of MHD test problems. We also describe Cholla's performance and weak scaling behavior on up to 74,088 GPUs using \textit{Frontier}. In Section \ref{sec:testing}, we discuss the new continuous integration and automated testing framework. We conclude in Section \ref{sec:summary}. In figure captions the Zenodo icon, \img{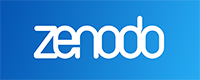}, links to the Zenodo repository that contains the python scripts to generate each figure. This Zenodo repository has sufficient information to reproduce the figures shown in this paper.

\section{Methods}
\label{sec:methods}



\subsection{Magnetohydrodynamics}
\label{sec:methods-mhd}

Cholla solves the ideal MHD equations in their conserved Eulerian form using a finite volume method \citep{Godunov}. These equations neglect all dissipative processes, including finite viscosity, electrical resistivity, and thermal conductivity. These approximations are reasonable when simulating regions of with very high Reynolds numbers as is common in many astrophysical problems. We note that although we neglect these additional processes at present, the methods implemented here are fully compatible with future extensions to include additional physics that depends on magnetic fields, such as anisotropic conduction, cosmic ray transport, or non-ideal MHD effects.

The ideal MHD equations are:

\begin{equation}
    \label{eqn:mass-conservation}
    \frac{\partial \rho}{\partial t} + \nabla \cdot (\rho \boldsymbol{v}) = 0
\end{equation}

\begin{equation}
    \label{eqn:momentum-conservation}
    \frac{\partial \rho\boldsymbol{v}}{\partial t} + \nabla \cdot (\rho \boldsymbol{v}\otimes\boldsymbol{v} - \boldsymbol{B}\otimes\boldsymbol{B} + P_T\boldsymbol{I}) = 0
\end{equation}

\begin{equation}
    \label{eqn:energy-conservation}
    \frac{\partial E}{\partial t} + \nabla \cdot ( (E + P_T) \boldsymbol{v} + \boldsymbol{B}(\boldsymbol{B}\cdot\boldsymbol{v}) ) = 0
\end{equation}

\begin{equation}
    \label{eqn:induction}
    \frac{\partial \boldsymbol{B}}{\partial t} - \nabla \times (\boldsymbol{v} \times \boldsymbol{B}) = 0
\end{equation}

\noindent where $\rho$ is density, $\boldsymbol{v} = ( v_x, v_y, v_z)$ is the velocity vector, $t$ is time, $\boldsymbol{B} = ( B_x, B_y, B_z)$ is the magnetic field, $\boldsymbol{I}$ is the identity tensor, $P_T \equiv P + \frac{1}{2}(\boldsymbol{B} \cdot \boldsymbol{B})$ is the total pressure, and $E$ is the total energy per unit volume $E \equiv \epsilon + \frac{1}{2}\rho(\boldsymbol{v}\cdot\boldsymbol{v}) + \frac{1}{2}(\boldsymbol{B}\cdot\boldsymbol{B})$. We adopt units in which the magnetic permeability $\mu_0 = 1$.

Equation \ref{eqn:mass-conservation} describes the conservation of mass, equation \ref{eqn:momentum-conservation} describes the conservation of momentum, equation \ref{eqn:energy-conservation} describes the conservation of energy, and equation \ref{eqn:induction} is the induction equation, which describes the divergence free condition. Cholla uses an ideal gas equation of state  which is $P \equiv \epsilon(\gamma - 1)$ where $\gamma$ is the adiabatic index and $\epsilon$ is the internal energy density.

In practice these equations are used in their vector form, where $\boldsymbol{U}$ and $\boldsymbol{W}$ are the conserved and primitive variables respectively:

\begin{align}
    \boldsymbol{U} &= \begin{bmatrix}
            \rho , &
            \rho v_x, &
            \rho v_y, &
            \rho v_z, &
            E,   &
            B_x, &
            B_y, &
            B_z
         \end{bmatrix}
    \\
    \boldsymbol{W} &= \begin{bmatrix}
            \rho, &
            v_x, &
            v_y, &
            v_z, &
            P,   &
            B_x, &
            B_y, &
            B_z,
         \end{bmatrix}.
\end{align}
The conservation equations, in Cartesian coordinates, can then be rewritten as

\begin{equation}
    \label{eqn:vector-conserved}
    \frac{\partial \boldsymbol{U}}{\partial t} +
    \frac{\partial \boldsymbol{F_x}}{\partial x} +
    \frac{\partial \boldsymbol{F_y}}{\partial y} +
    \frac{\partial \boldsymbol{F_z}}{\partial z} = 0
\end{equation}
where $\boldsymbol{F_x}$, $\boldsymbol{F_y}$, and $\boldsymbol{F_z}$ are the vectors of fluxes in the $x$, $y$, and $z$ direction respectively and are given by

\begin{equation}
    \boldsymbol{F_x} = \begin{bmatrix}
            \rho v_{x} \\
            \rho v_{x}^2 + P_{T} - B_{x}^2 \\
            \rho v_{x} v_{y} - B_{x} B_{y} \\
            \rho v_{x} v_{z} - B_{x} B_{z} \\
            v_{x} \left( E + p_{T} \right) - B_{x} \left( \boldsymbol{v} \cdot \boldsymbol{B} \right) \\
            0 \\
            B_{y} v_{x} - B_{x} v_{y} \\
            B_{z} v_{x} - B_{x} v_{z} \\
         \end{bmatrix}
\end{equation}

\begin{equation}
    \boldsymbol{F_y} = \begin{bmatrix}
            \rho v_{y} \\
            \rho v_{y} v_{x} - B_{y} B_{x} \\
            \rho v_{y}^2 + P_{T} - B_{y}^2 \\
            \rho v_{y} v_{z} - B_{y} B_{z} \\
            v_{y} \left( E + P_{T} \right) - B_{y} \left( \boldsymbol{v} \cdot \boldsymbol{B} \right) \\
            B_{x} v_{y} - B_{y} v_{x} \\
            0 \\
            B_{z} v_{y} - B_{y} v_{z} \\
         \end{bmatrix}
\end{equation}

\begin{equation}
    \boldsymbol{F_z} = \begin{bmatrix}
            \rho v_{z} \\
            \rho v_{z} v_{x} - B_{z} B_{x} \\
            \rho v_{z} v_{y} - B_{z} B_{y} \\
            \rho v_{z}^2 + P_{T} - B_{z}^2 \\
            v_{z} \left( E + P_{T} \right) - B_{z} \left( \boldsymbol{v} \cdot \boldsymbol{B} \right) \\
            B_{x} v_{z} - B_{z} v_{x} \\
            B_{y} v_{z} - B_{z} v_{y} \\
            0
         \end{bmatrix}.
\end{equation}

\subsection{The VL+CT Integrator}
\label{sec:vlct-summary}

To integrate these equations in Cholla, we implement the VL+CT (Van Leer plus Constrained Transport) integrator introduced in \cite{stone_2009}, along with the HLLD Riemann solver described in \cite{hlld_2005}. Many of the equations were first described in the context of the constrained-transport extension to the corner-transport upwind (CTU) method \citep{colella_1990} described in 2D \citep{gardiner_2005} and 3D \citep{gardiner_unsplit_2008} for the Athena MHD code \citep{stone_athena_2008}. Our specific implementation of the piecewise parabolic method follows \cite{felker_2018}, which is itself an extension of the original PPM method presented by \cite{colella_1984}. The VL+CT integrator is very similar in structure to the MUSCL-Hancock integrator \citep{van_leer_2006, falle_1991, Toro}, though with significant new additions for Constrained Transport (CT). We note that this is now the default integrator used in Cholla, as we have found it to be more robust than the CTU algorithm described in \cite{schneider_2015}.

Constrained transport treats the magnetic field as a surface averaged quantity centered at cell interfaces rather than a volume averaged, cell centered quantity like the hydro variables; i.e. a staggered grid. Each face stores only the magnetic field perpendicular to that face, i.e. the $x,i+1/2,j,k$ face stores the $B_{x,i+1/2,j,k}$ magnetic field. At each time step, the magnetic field is then updated using edge averaged electric fields computed from the magnetic flux returned by the Riemann solver. Updating the magnetic field with the electric field automatically fulfills the divergence free condition for magnetic fields, assuming that the initial conditions are also divergence free.

A brief overview of the algorithm for a single time step using the VL+CT integrator is as follows; more detailed discussion of each step is presented in the following subsections.

\begin{enumerate}
    \item Compute the time step, $\Delta t$.
    \item Reconstruct interface states using a piecewise constant method (PCM) approximation.
    \item Solve the Riemann problem on every interface using the PCM interface states.
    \item Compute the edge centered electric fields.
    \item Update the $t=n$ conserved variables to $t=n+\Delta t/2$ using the computed fluxes and electric fields.
    \item Reconstruct interface states using a piecewise linear method (PLM) or piecewise parabolic method (PPM) approximation.
    \item Solve the Riemann problem on every interface using the higher order reconstructed interface states.
    \item Recompute the edge centered electric fields.
    \item Update the $t=n$ conserved variables to $t=n+\Delta t$ using the half time step fluxes and electric fields
\end{enumerate}


\subsubsection{Step 1: Compute the Time Step}
\label{vlct:dt}

The first step is to compute the time step. Cholla implements uniform time steps across the grid, so this is the minimum crossing time of any wave in any cell multiplied by a Courant factor to maintain the Courant–Friedrichs–Lewy condition \cite{cfl}:

\begin{equation}
        \label{eqn:dt}
        \Delta t = C_{CFL} \min \left(
            \frac{\Delta x}{\mid v^n_{x,i,j,k} \mid + c^n_{f,i,j,k}},
            \frac{\Delta y}{\mid v^n_{y,i,j,k} \mid + c^n_{f,i,j,k}},
            \frac{\Delta z}{\mid v^n_{z,i,j,k} \mid + c^n_{f,i,j,k}}
        \right),
\end{equation}

\noindent where $\Delta t$ is the time step, $C_{CFL} \leq 0.5$ is the CFL number, $\mid v^n_{l,i,j,k}\mid $ is the magnitude of the velocity in the $l$ direction velocity in the ${i,j,k}$ cell, and $c^n_f $ is the fast magnetosonic wave-speed computed using \emph{cell centered} values. The method is formally accurate at $C_{CFL} = 0.5$, though that is an estimate\citep{stone_2009}. We have found empirically that the method is not stable at $C_{CFL} = 0.5$ but is stable at $C_{CFL} = 0.4$\footnote{Typically we use $C_{CFL} = 0.3$ to improve the accuracy of the operator splitting}. The fast and slow magnetosonic speeds are

\begin{equation}
    c_{f,s} = \sqrt{\frac
    {\gamma p + \mid \boldsymbol{B} \mid^2 \pm \sqrt{\left( \gamma p \;+ \mid \boldsymbol{B} \mid^2 \right)^2 - 4\gamma p B_x^2 } }
    {2\rho}}
\end{equation}

\noindent where the $+$ option corresponds to the fast magnetosonic speed, and $-$ to the slow.

Equation \ref{eqn:dt} computes the minimum crossing time of a wave in a specific cell. A global reduction is then performed to find the minimum in the entire grid; that minimum is used as the time step $\Delta t$. This reduction is done primarily on the GPU followed by an MPI ALL\_REDUCE. Because GPUs programming methods are inherently parallel, GPU reductions are not trivial and their performance is sensitive to the method used. Details of our reduction method are discussed in section \ref{sec:gpu-vs-cpu}.

The cell-centered magnetic field is computed with a direct average of the face centered values:
\begin{equation}
    \begin{aligned}
        B^n_{x,i,j,k,} = \frac{1}{2} \left( B^n_{x,i+1/2,j,k} + B^n_{x,i-1/2,j,k} \right) \\
        B^n_{y,i,j,k,} = \frac{1}{2} \left( B^n_{y,i,j+1/2,k} + B^n_{y,i,j-1/2,k} \right) \\
        B^n_{z,i,j,k,} = \frac{1}{2} \left( B^n_{z,i,j,k+1/2} + B^n_{z,i,j,k-1/2} \right) \\
    \end{aligned}
\end{equation}
These cell-centered values for the magnetic field are used several times throughout the integrator. In CPU-based codes it is typically more efficient to compute these cell centered magnetic fields once and save them, since traditional CPU-based codes are typically compute limited. Cholla is generally limited instead by GPU memory, so we recompute the cell-centered values when they are needed rather than permanently allocating the associated memory.

\subsubsection{Step 2: First Order Reconstruction}
\label{vlct:first-order-reconstruction}

Reconstructing the interface states at first order, or the piecewise constant method (PCM), is done by setting the primitive interface states values to the same value as the cell:

\begin{equation}
    \boldsymbol{W}_{L, i+1/2} = \boldsymbol{W}_{R, i-1/2} = \boldsymbol{W}_{i}
\end{equation}

\noindent where $ \boldsymbol{W}_{L/R, i\pm1/2} $ is the state on the left or right side of the cell. Although Cholla evolves the conserved variables, primitive variables are typically used for interface reconstruction since they are required by the Riemann solver used to compute the fluxes. In higher order spatial reconstructions we also use the characteristic variables derived from the primitive variables.

Although PCM is too diffusive to be used on its own in both reconstruction steps, we note that that mode is excellent for debugging. Here, it offers a computationally-inexpensive way to calculate first-order fluxes, which will be used for the ``predictor" step of this predictor-corrector algorithm\citep{van_leer_2006, stone_2009}. Most higher order methods revert to PCM near large discontinuities\citep{leveque2002finite}.

\subsubsection{Step 3: First Riemann Solve}
\label{vlct:first-riemann-solve}

The next step is to solve the Riemann problem with the first order interface states. To do this, we employ the HLLD Riemann solver introduced in \cite{hlld_2005}. Because we have implemented the Riemann solver exactly as described in \cite{hlld_2005} we do not reproduce all the details here. The longitudinal magnetic field does not require reconstruction and can be used directly since it is stored at the face. The transverse fields (i.e. the fields parallel to the interface) are reconstructed from the cell centered average (computed as shown in section \ref{vlct:dt}), then reconstructed using PCM identically to other fields.

The magnetic fluxes returned by the Riemann solver are the face centered electric fields \citep[see section 5.3 of][]{stone_athena_2008}. For clarity, we list in Table \ref{table:emf} the correlation between flux and electric field.

\begin{deluxetable*}{llll}
    \label{table:emf}
    \tablecaption{Magnetic Flux to Face Centered electric field}

    \tablehead{\colhead{HLLD Solve Direction} & \colhead{Equation for Magnetic Flux} & \colhead{Eqn. as a Cross Product} & \colhead{electric field}}

    \startdata
    $ X $ & $ V_x B_y - B_x V_y $ & $  (V \times B)_z $ & $ -\varepsilon_z $ \\ \hline
    $ X $ & $ V_x B_z - B_x V_z $ & $ -(V \times B)_y $ & $  \varepsilon_y $ \\ \hline
    $ Y $ & $ V_x B_y - B_x V_y $ & $  (V \times B)_z $ & $ -\varepsilon_x $ \\ \hline
    $ Y $ & $ V_x B_z - B_x V_z $ & $ -(V \times B)_y $ & $  \varepsilon_z $ \\ \hline
    $ Z $ & $ V_x B_y - B_x V_y $ & $  (V \times B)_z $ & $ -\varepsilon_y $ \\ \hline
    $ Z $ & $ V_x B_z - B_x V_z $ & $ -(V \times B)_y $ & $  \varepsilon_x $ \\ \hline
    \enddata

    \tablecomments{The directions used here are relative to the internal workings of the HLLD solver. Since the HLLD solver is inherently 1D we run it once for each of the faces of a cell. So in the case where the solver is running in the Y direction the solver's Y field is actually the Z field and the solver's Z field is actually the X field, cyclically extended for the Z direction.}
\end{deluxetable*}

\subsubsection{Step 4: Compute the Constrained Transport Electric Field}
\label{vlct:emf}

The next step is to calculate the constrained transport electric field. These fields are \emph{line averaged} along each cell vertex. These line averaged fields are constructed by averaging the face centered electric fields from the previous step, and their slopes. Equation \ref{eqn:emf-edge} is used to reconstruct these line averaged electric fields; only the equation for the $ z $-direction is presented, the equations for the other directions can be found by cyclic permutation. The equations for computing the other directions are obtained by substituting out the $ z $ index with $ x $ or $ y $ and changing the derivatives appropriately.

\begin{equation}
    \label{eqn:emf-edge}
    \begin{aligned}
        \mathcal{E}_{z, i-1/2, j-1/2, k} = \frac{1}{4} \left(
              \mathcal{E}_{z, i-1/2, j, k}
            + \mathcal{E}_{z, i, j-1/2, k}
            + \mathcal{E}_{z, i-1/2, j-1, k}
            + \mathcal{E}_{z, i-1, j-1/2, k}\right) \\
        + \frac{\Delta y}{8} \left( \left( \frac{\partial \mathcal{E}_z }{\partial y} \right)_{i-1/2, j-1/4, k} + \left(  \frac{\partial \mathcal{E}_z }{\partial y} \right)_{i-1/2, j-3/4, k} \right) \\
        + \frac{\Delta x}{8} \left( \left( \frac{\partial \mathcal{E}_z }{\partial x} \right)_{i-1/4, j-1/2, k} + \left(  \frac{\partial \mathcal{E}_z }{\partial x} \right)_{i-3/4, j-1/2, k} \right)
    \end{aligned}
\end{equation}

\begin{figure}[ht!]
    \includegraphics[width=0.5\linewidth]{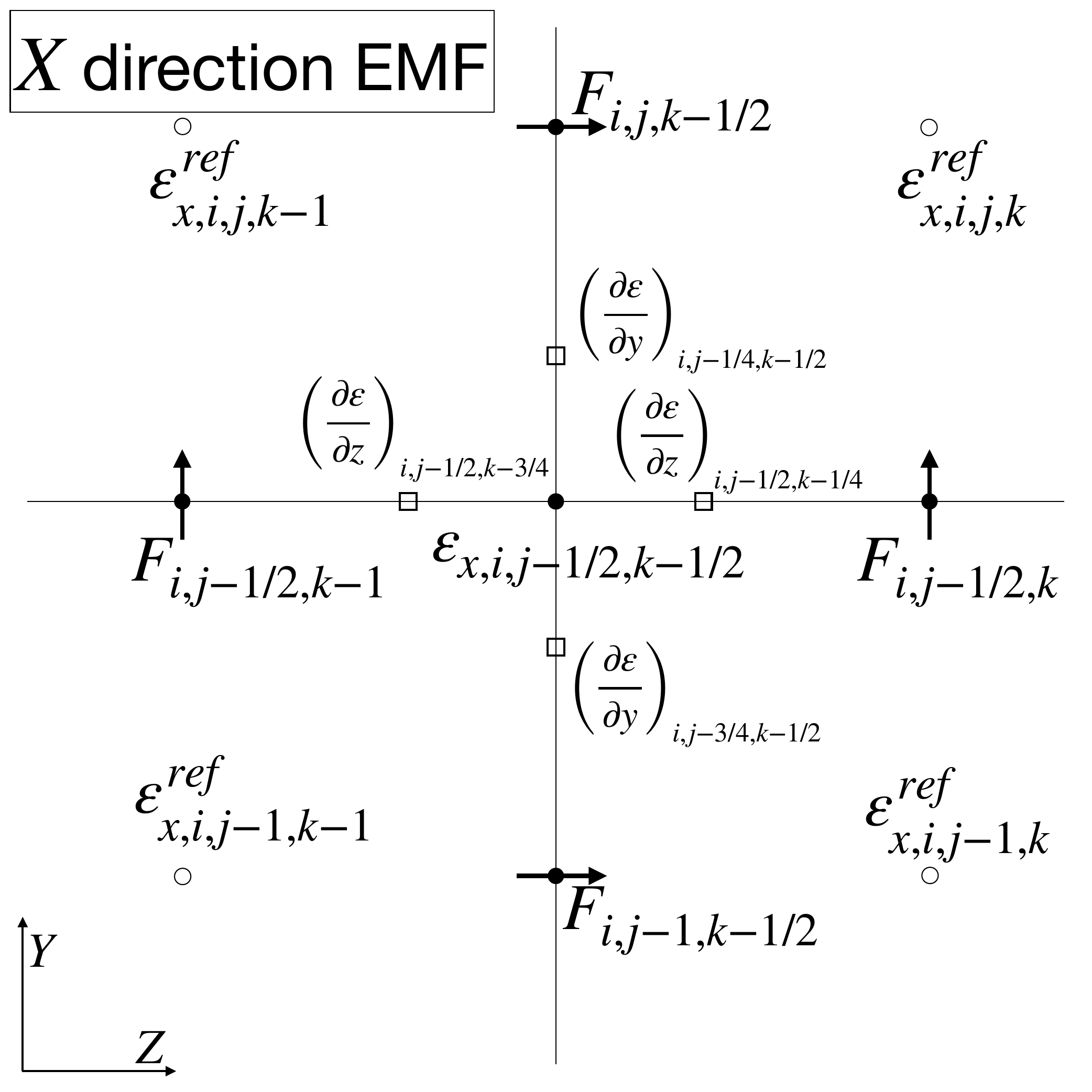}
    \includegraphics[width=0.5\linewidth]{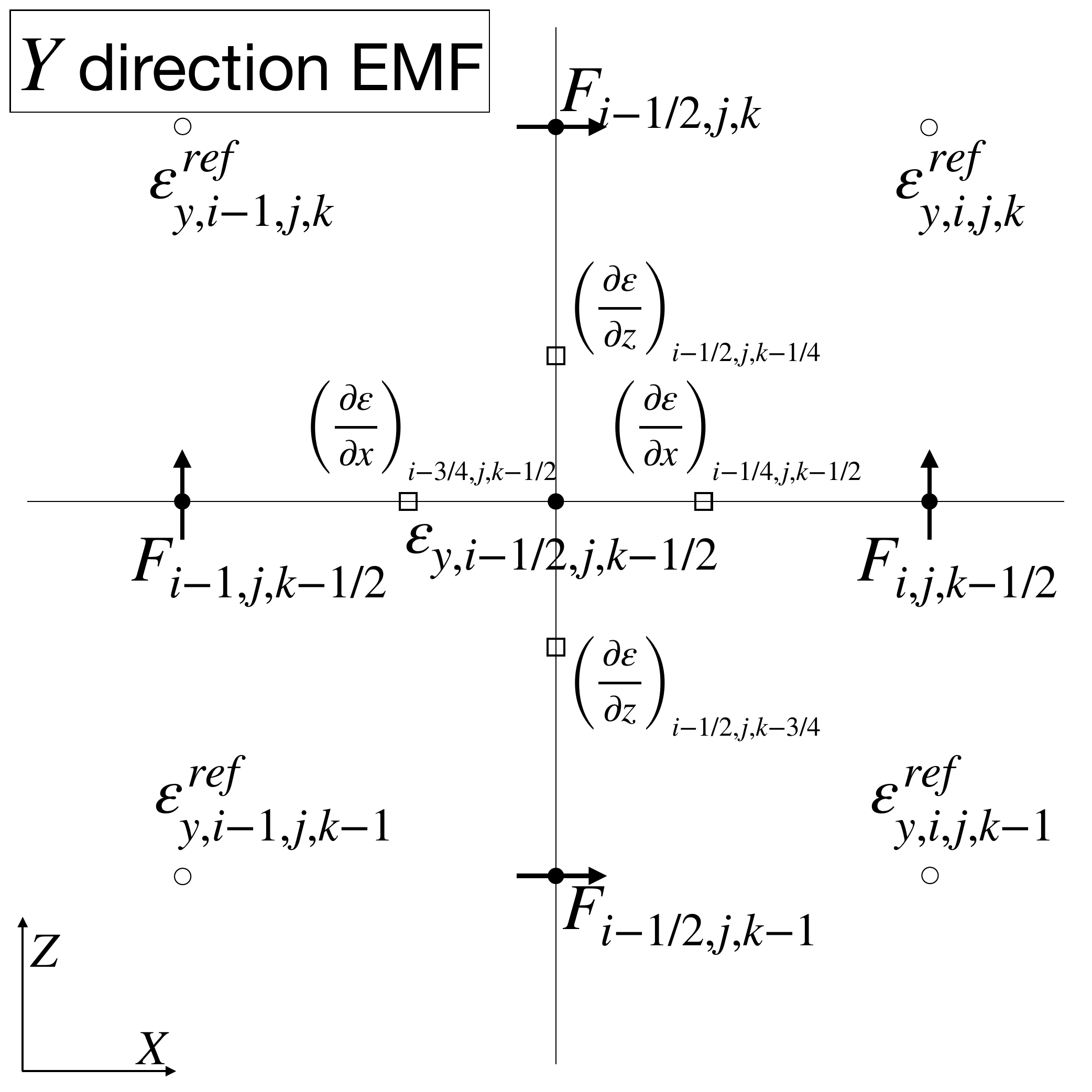}
    \includegraphics[width=0.5\linewidth]{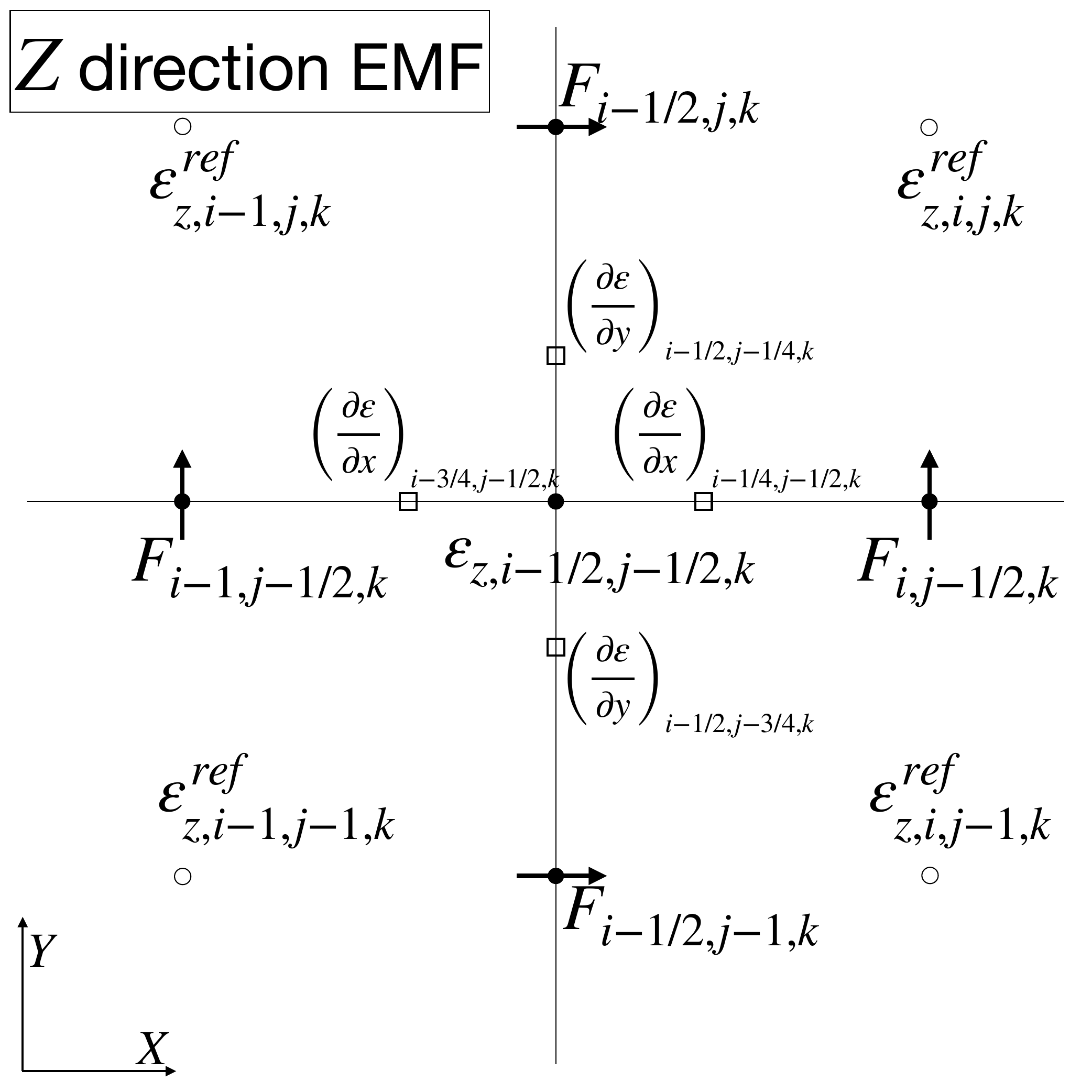}
    \caption{2D slices in all three planes showing the location of the fluxes, edge electric fields, and derivatives. Based on Figure 5 of \cite{stone_athena_2008}.}
    \label{fig:emf-graph}
\end{figure}

$\mathcal{E}_{z, i-1/2, j-1/2, k}$ is the line averaged electric field; the first four terms on the right are the face averaged electric fields, and the four derivative terms are the derivatives of those fields in the direction towards the edge, details of which are in Equation \ref{eqn:emf-slope}. Figure \ref{fig:emf-graph} shows the spatial relationships between the derivatives and the relevant edge states and reference fields. Each edge requires 4 derivatives and they are computed as differences between a reference state and an edge state.

Technically, constrained transport uses the magnetic flux as the conserved variable and EMF (ElectroMotive Force) to update it. However, because those values only differ from the magnetic flux density (i.e. the magnetic field) and the electric field respectively by factors of unit length either can be treated as the conserved variable, the same way either density or mass can be evolved in the hydro fields \citep{stone_athena_2008}. As such we use the magnetic field and electric field to evolve the grid. Electric fields have the proper units to evolve the magnetic flux density, $ B $, whereas EMF has the proper units to evolve the magnetic flux.

On any face there are two non-zero electric fields; both transverse to the face. The component that is used to calculate the field along a given edge is the component that is parallel to that edge; i.e. if the edge points along the $ z $-direction then the field pointing along the $ z $-direction is used, not the field in the $ x $ or $ y $ direction.

The derivatives from Equation \ref{eqn:emf-edge} are computed using the upwinded slope as follows

\begin{equation}
    \label{eqn:emf-upwind-slope}
    \left( \frac{\partial \mathcal{E}_z }{\partial y} \right)_{i-1/2, j-1/4, k} =
        \begin{cases}
            \left( \frac{\partial \mathcal{E}_z }{\partial y} \right)_{i-1, j-1/4, k} & \text{for} \; v_{x, i-1/2} > 0
            \\
            \\
            \left( \frac{\partial \mathcal{E}_z }{\partial y} \right)_{i, j-1/4, k} & \text{for} \; v_{x, i-1/2} < 0
            \\
            \\
            \frac{1}{2} \left( \left( \frac{\partial \mathcal{E}_z }{\partial y} \right)_{i-1, j-1/4, k} + \left( \frac{\partial \mathcal{E}_z }{\partial y} \right)_{i, j-1/4, k} \right) & \text{otherwise}
            \\
            \\
        \end{cases}
\end{equation}

\noindent where, for example, the derivatives are given by

\begin{equation}
    \label{eqn:emf-slope}
    \left( \frac{\partial \mathcal{E}_z }{\partial y} \right)_{i, j-1/4, k} =
    2 \left( \frac{\mathcal{E}_{z,i,j-1/2,k} - \mathcal{E}_{z,i,j,k}^{ref}}{\Delta y} \right).
\end{equation}

\noindent $\mathcal{E}_{z,i,j,k}^{ref}$, is the cell centered reference field. This reference field is computed with the following cross product

\begin{equation}
    \mathcal{E}_{i,j,k}^{ref,n} = - \left( \boldsymbol{v}^{n}_{i.j.k} \times \boldsymbol{B}^{n}_{i.j.k} \right).
\end{equation}

\subsubsection{Step 5. Perform the Half Time-step Update}
\label{vlct:half-dt-update}

We first update the density, momenta, and energy, but not the magnetic fields, using the standard conservative update equation and the first-order fluxes from Step 3:

\begin{equation}
    \begin{aligned}
        \boldsymbol{U}^{n+1/2}_{i,j,k} = \boldsymbol{U}^{n}_{i,j,k}
        - \frac{\Delta t}{\Delta x} \left( \boldsymbol{F}^n_{x,i+1/2,j,k} - \boldsymbol{F}^n_{x,i-1/2,j,k} \right) \\
        - \frac{\Delta t}{\Delta y} \left( \boldsymbol{F}^n_{y,i,j+1/2,k} - \boldsymbol{F}^n_{y,i,j+1/2,k} \right) \\
        - \frac{\Delta t}{\Delta z} \left( \boldsymbol{F}^n_{z,i,j,k+1/2} - \boldsymbol{F}^n_{z,i,j,k+1/2} \right).
    \end{aligned}
\end{equation}

We then update the magnetic field using the electric fields computed in Step 4:

\begin{equation}
    \begin{aligned}
        B^{n+1/2}_{x,i-1/2,j,k} = B^{n}_{x,i-1/2,j,k}
        + \frac{\Delta t}{\Delta z} \left( \mathcal{E}^n_{y,i-1/2,j,k+1/2} - \mathcal{E}^n_{y,i-1/2,j,k-1/2} \right) \\
        - \frac{\Delta t}{\Delta y} \left( \mathcal{E}^n_{z,i-1/2,j+1/2,k} - \mathcal{E}^n_{z,i-1/2,j-1/2,k} \right)
    \end{aligned}
\end{equation}

\begin{equation}
    \begin{aligned}
        B^{n+1/2}_{y,i,j-1/2,k} = B^{n}_{y,i,j-1/2,k}
        + \frac{\Delta t}{\Delta x} \left( \mathcal{E}^n_{z,i+1/2,j-1/2,k} - \mathcal{E}^n_{z,i-1/2,j-1/2,k} \right) \\
        - \frac{\Delta t}{\Delta z} \left( \mathcal{E}^n_{x,i,j-1/2,k+1/2} - \mathcal{E}^n_{x,i,j-1/2,k-1/2} \right)
    \end{aligned}
\end{equation}

\begin{equation}
    \begin{aligned}
        B^{n+1/2}_{z,i,j,k-1/2} = B^{n}_{z,i-1/2,j,k}
        + \frac{\Delta t}{\Delta y} \left( \mathcal{E}^n_{x,i,j+1/2,k-1/2} - \mathcal{E}^n_{x,i,j-1/2,k-1/2} \right) \\
        - \frac{\Delta t}{\Delta x} \left( \mathcal{E}^n_{y,i+1/2,j,k-1/2} - \mathcal{E}^n_{y,i-1/2,j,k-1/2} \right).
    \end{aligned}
\end{equation}

\subsubsection{Step 6. Half Time-step Second Order Reconstruction}
\label{vlct:higher-order-reconstruction}

Step 5 results in first order time-averaged values for the cell-centered conserved variables and face-centered magnetic fields. To make the integration second-order in time, we need to perform a ``corrector" step. First, we perform a higher order interface reconstruction. The method shown here is for Piecewise Linear Method (PLM), reconstruction. Cholla currently implements piecewise constant, piecewise linear, and piecewise parabolic reconstruction, with limiting in the characteristic variables, for MHD. The piecewise parabolic method that Cholla utilizes is discussed in detail in \cite{felker_2018}. Using the third order piecewise parabolic method for spatial reconstruction does typically give slightly more accurate results at a given resolution compared to PLM, but since the method is formally second order it does not improve the overall order of convergence. Note that at a given face only the transverse components of the electric field need to be reconstructed. The longitudinal component is already given at the face.

The steps of the PLM update are:
\begin{enumerate}
    \item Compute the primitive variables from the conserved variables.

    \item Compute the left, right, centered, and Van Leer differences in the primitive variables
    \begin{align}
        \delta \boldsymbol{W}_{L,i} &= \boldsymbol{W_{i}} - \boldsymbol{W_{i-1}} \\
        \delta \boldsymbol{W}_{R,i} &= \boldsymbol{W_{i+1}} - \boldsymbol{W_{i}} \\
        \delta \boldsymbol{W}_{C,i} &= \frac{\boldsymbol{W_{i+1}} - \boldsymbol{W_{i-1}}}{2} \\
        \delta \boldsymbol{W}_{VL,i} &=
        \begin{cases}
            \frac{2 \boldsymbol{W}_{L,i} \boldsymbol{W}_{R,i}}{\boldsymbol{W}_{L,i} +\boldsymbol{W}_{R,i}} ,& \text{if } \boldsymbol{W}_{L,i} \boldsymbol{W}_{R,i} > 0\\
            0,              & \text{otherwise}
        \end{cases}
    \end{align}

    \item Project the slopes into the characteristic variables, $\boldsymbol{a}$, using the eigenvectors listed in the appendix of \cite{stone_athena_2008}. Note that to maintain mathematical consistency we use the eigenvectors of the $\boldsymbol{w_{i}}$ cell for all four slopes and the later projection back to primitive variables.
    \item Apply monotonicity constraints to the characteristic differences to ensure that the reconstruction is total variation diminishing (TVD). We use the following limiter, given in \cite{leveque2002finite}:
    \begin{equation}
        \label{eqn:limiter}
        \delta \boldsymbol{a}_{m,i} =
        \begin{cases}
             \text{sgn}(\boldsymbol{a}_{C,i})\min(2\abs{\boldsymbol{a}_{L,i}},2\abs{\boldsymbol{a}_{R,i}},\abs{\boldsymbol{a}_{C,i}},\abs{\boldsymbol{a}_{VL,i}}),& \text{if } \boldsymbol{a}_{L,i} \boldsymbol{a}_{R,i} > 0\\
            0,              & \text{otherwise}
        \end{cases}
    \end{equation}
    \item Project the limited characteristic slopes, $\delta \boldsymbol{a}_{m}$, back into the primitive variables, $\delta \boldsymbol{W}_{m}$, using the eigenvectors.
    \item Compute the interface states $\boldsymbol{W}_{L, i+1/2}$ and $\boldsymbol{W}_{R, i-1/2}$:
\end{enumerate}

    \begin{equation}
        \begin{aligned}
            \boldsymbol{W}_{L, i+1/2} = \boldsymbol{W}_{i} + \frac{\delta \boldsymbol{W}_{m, i}}{2} \\
            \boldsymbol{W}_{R, i-1/2} = \boldsymbol{W}_{i} - \frac{\delta \boldsymbol{W}_{m, i}}{2} \\
        \end{aligned}
    \end{equation}

    \noindent where $ \boldsymbol{W}_{L/R, i\pm1/2} $ is the state on the left or right side
    of the cell and $ \delta \boldsymbol{W}_{m, i} $ is the monotonically limited
    primitive slope. Limiting can be done in the primitive variables by skipping steps 3 and 5 and replacing the characteristic slopes in Equation \ref{eqn:limiter} with their primitive counterparts.

\subsubsection{Step 7. Second Riemann Solve}
\label{vlct:2nd-riemann-solve}

Solve the Riemann problem for each interface again using the higher order interface states computed in step 6.

\subsubsection{Step 8. Compute the Constrained Transport Electric Fields}
\label{vlct:2nd-emf}

Repeat step 3, but using the fluxes from the second Riemann solve and the half time step MHD variables computed in Step 5.

\subsubsection{Step 9. Perform the Full Time-step Update}
\label{vlct:full-dt-update}

Update the cell-centered hydro variables from $t = n$ to $t = n + \Delta t$ using the conserved update equation and the second order fluxes computed in Step 7:

\begin{equation}
    \begin{aligned}
        \boldsymbol{U}^{n+1}_{i,j,k} = \boldsymbol{U}^{n}_{i,j,k}
        &- \frac{\Delta t}{\Delta x} \left( \boldsymbol{F}^{n+1/2}_{x,i+1/2,j,k} - \boldsymbol{F}^{n+1/2}_{x,i-1/2,j,k} \right) \\
        &- \frac{\Delta t}{\Delta y} \left( \boldsymbol{F}^{n+1/2}_{y,i,j+1/2,k} - \boldsymbol{F}^{n+1/2}_{y,i,j+1/2,k} \right) \\
        &- \frac{\Delta t}{\Delta z} \left( \boldsymbol{F}^{n+1/2}_{z,i,j,k+1/2} - \boldsymbol{F}^{n+1/2}_{z,i,j,k+1/2} \right).
    \end{aligned}
\end{equation}

Update the face-centered magnetic field using the electric fields calculated in Step 8:

\begin{equation}
    \begin{aligned}
        B^{n+1}_{x,i-1/2,j,k} = B^{n}_{x,i-1/2,j,k}
        + \frac{\Delta t}{\Delta z} \left( \mathcal{E}^{n+1/2}_{y,i-1/2,j,k+1/2} - \mathcal{E}^{n+1/2}_{y,i-1/2,j,k-1/2} \right) \\
        - \frac{\Delta t}{\Delta y} \left( \mathcal{E}^{n+1/2}_{z,i-1/2,j+1/2,k} - \mathcal{E}^{n+1/2}_{z,i-1/2,j-1/2,k} \right)
    \end{aligned}
\end{equation}

\begin{equation}
    \begin{aligned}
        B^{n+1}_{y,i,j-1/2,k} = B^{n}_{y,i,j-1/2,k}
        + \frac{\Delta t}{\Delta x} \left( \mathcal{E}^{n+1/2}_{z,i+1/2,j-1/2,k} - \mathcal{E}^{n+1/2}_{z,i-1/2,j-1/2,k} \right) \\
        - \frac{\Delta t}{\Delta z} \left( \mathcal{E}^{n+1/2}_{x,i,j-1/2,k+1/2} - \mathcal{E}^{n+1/2}_{x,i,j-1/2,k-1/2} \right)
    \end{aligned}
\end{equation}

\begin{equation}
    \begin{aligned}
        B^{n+1}_{z,i-1/2,j,k} = B^{n}_{z,i-1/2,j,k}
        + \frac{\Delta t}{\Delta y} \left( \mathcal{E}^{n+1/2}_{x,i,j+1/2,k-1/2} - \mathcal{E}^{n+1/2}_{x,i,j-1/2,k-1/2} \right) \\
        - \frac{\Delta t}{\Delta x} \left( \mathcal{E}^{n+1/2}_{y,i+1/2,j,k-1/2} - \mathcal{E}^{n+1/2}_{y,i-1/2,j,k-1/2} \right).
    \end{aligned}
\end{equation}

\subsubsection{Step 10. Increment the Time by \texorpdfstring{$\Delta t$}{dt}}
\label{vlct:increment-time}

Increment the time by $\Delta t$. Additional physics modules are added in a first order operator-split fashion after this point (chemistry, radiation transport, etc.), after which all MPI communication is done to exchange the ghost/halo cells that surround each MPI rank's subdomain.

\subsection{Implementation on GPUs}
\label{sec:gpu-vs-cpu}

\subsubsection{Memory bandwidth constraints}
While the implementation of MHD on GPUs is similar to the implementation on CPUs, there are some crucial differences, especially regarding data handling and movement. Compared to CPUs, GPUs have higher memory bandwidth and extremely high FLOPS, but limited memory capacity and limited functionality due to their fundamentally SIMD (Single Instruction Multiple Data) nature. Also, while GPU memory bandwidth is higher on the whole, due to their parallel nature GPUs can request many more values from memory at once, meaning GPU-based codes are often still memory bandwidth bound.

This leads to some implementation choices that may seem counter-intuitive. For example, while an optimized CPU-based code may compute the cell-centered magnetic fields once and then save them to memory, Cholla recomputes them in each function call where they are required. Similarly, Cholla does not store the primitive variables, only the conserved ones, and recomputes the primitive variables as needed. This approach reduces global memory usage by not storing an entire second grid. It also generally reduces memory bandwidth requirements, since often both the conserved and primitive variables are needed within a function, but only one set needs to be loaded.

Another potential source of memory bandwidth optimization is fusing functions together. By combining multiple GPU kernels into one, different functions can share data without a write/read cycle to global memory. For example, we have implemented this in Cholla by combining the first order reconstruction and Riemann solver kernels, since each interface that is reconstructed is only used in the next Riemann solve, and does not need to be used later in the integrator. Fusing the PCM ``reconstruction" into the Riemann solver led to a $\sim 10\%$ improvement in overall runtime. However, fusing the PLMC reconstruction into the Riemann solver actually slowed the code down by $\sim 10-15\%$. This slowdown is caused by the increased register usage of the new, larger kernel, which in turn reduced overall GPU occupancy. Thus, the trade-off between reducing global memory accesses and increasing register usage is not always straightforward to determine, and we have found that it depends on compiler optimizations which can vary between platforms, as well.

Another major challenge in GPU computing is CPU-to-GPU bandwidth which is much lower than GPU memory bandwidth. To address this problem, Cholla now keeps as much data as possible in the GPU memory instead of moving it to and from main system memory, in contrast with previous versions of the code. Although it has always been a GPU-native code, historically Cholla used an extreme version of the ``offload" model of GPU programming, keeping the ``primary" data (like conserved variable arrays) in CPU memory, and copying that data to the GPU to carry out computations before copying it back each time step. Although this approach originally had the advantage of allowing larger grid sizes to be computed on a single GPU when GPU memory sizes were extremely limited, as GPUs have gotten dramatically faster and their memory capacity has increased over the last decade, these full-grid copies between CPU and GPU memory would now dominate the simulation run time. In addition, most new supercomputers, such as \textit{Frontier}, have GPUs that are directly connected to the network, so direct off-node GPU-to-GPU MPI communication is now possible. Between these two factors it is much more efficient to store all the simulation data on the GPU, and only move it back to the CPU for i/o operations. In the new version of Cholla described in this work, all computations are carried out on the GPU, and  MPI communication proceeds directly from GPU memory buffers. The only functions that remain on the CPU are the setting of the initial conditions, which only occurs once, and reading and writing output files.\footnote{At the time of writing, the HDF5 library does not support writing files directly from GPU memory.}

\subsubsection{Performance portability}

Cholla is written in C++ and CUDA, the C++ extension developed by NVIDIA. However, CUDA code can only be compiled using NVIDIA's compiler \texttt{nvcc} for NVIDIA GPUs. Thus, with the advent of AMD-based supercomputers like \emph{Frontier}, the problem of code portability becomes relevant. We approached this problem in two ways. First, AMD has developed HIP, a cross platform equivalent to CUDA, which can be compiled using their \texttt{hipcc} to target either NVIDIA or AMD hardware. Although porting Cholla's CUDA code to HIP did not prove challenging due to the extreme similarity between the two platforms, rewriting the entire code base in HIP has the drawback of making it impossible to compile on systems with NVIDIA GPUs that do not have HIP installed. Thus, we chose to employ a less invasive modification: we introduced a header file (\texttt{gpu.hpp}) which uses C++ preprocessor macros to convert between CUDA and HIP versions of functions at compile time. When building the code, the user merely has to specify whether the compilation should use \texttt{hipcc} or \texttt{nvcc}, and the code is compiled into HIP or CUDA accordingly. This method also has the advantage of continuing to allow us to maintain a single code-base written in CUDA.

\subsubsection{GPU reductions}

There are several places in Cholla where a grid wide reduction must be performed. The primary example is the calculation of the time step, which requires finding the minimum crossing time in the full simulation grid. While CPU and MPI based reductions are relatively simple to implement, often through library calls, the inherently parallel nature of the GPU programming model makes GPU based reductions somewhat more complex. For GPU reductions in Cholla, we use a method similar to that described by NVIDIA\footnote{\url{https://developer.nvidia.com/blog/faster-parallel-reductions-kepler/}} which describe the challenges of a GPU reduction well. This reduction method uses atomics for the final level of reduction. While this method is straightforward to implement in HIP, at the time of writing, CUDA does not support floating point \texttt{atomicMax}. To work around this we have adopted a method from the RAPIDS cuML library\footnote{\url{https://github.com/rapidsai/cuml/blob/dc14361ba11c41f7a4e1e6a3625bbadd0f52daf7/cpp/src\_prims/stats/minmax.cuh}} which, with some encoding, uses the integral \texttt{atomicMax}. Overall this method performs slightly faster than Cholla's previous hybrid GPU+CPU reduction for the number of elements we typically encounter. Primarily though, it is much simpler to use and performs dramatically better as the number of elements increases.

\section{MHD Tests}
\label{sec:mhd-tests}

In this Section, we show the results of a suite of test problems that demonstrate the accuracy, robustness and performance of Cholla MHD. We have included many problems that are common in the literature and specifically selected tests that are challenging for the integrator and, when possible, have quantifiable measures of correctness for comparison to other methods and codes. All of these problems are also included in our automated test suite as either accuracy and/or regression system tests (see Section \ref{sec:testing} for details). Section \ref{sec:accuracy_tests} discusses tests for accuracy, while Section \ref{sec:mhd-perf-tests} discusses the performance and scaling of Cholla MHD.

\subsection{Accuracy Tests}
\label{sec:accuracy_tests}

\subsubsection{Linear Wave Convergence}
\label{sec:lwc}

The propagation of the four MHD linear waves provides an excellent quantitative measure of the accuracy of a numberical MHD method. Our linear wave tests use a periodic domain of $1.0\times1.0\times1.0$ and a resolution of $N\times16\times16$ where $N$ goes from 16 to 512 in powers of 2. The wave equation is

\begin{equation}
    q = \overline{q} + A R_w \sin{\frac{2\pi x}{\lambda}},
\end{equation}

\noindent where $q$ is the conserved variable, $\overline{q}$ is the mean background state, $A=10^{-6}$ is the amplitude of the wave, $R_w$ is the right eigenvector in conserved variables for the wave mode $w$, $x$ is the position, and $\lambda=1$ is the wavelength of the wave. The adiabatic index $\gamma$ is $5/3$ and the background state is:
$\overline{\rho}=1.0$,
$\overline{v_x}=\overline{v_y}=\overline{v_z}=0$ (except for the contact wave where $\overline{v_x} = 1$),
$\overline{P}=1/\gamma$,
$\overline{B_x}=1$,
$\overline{B_y}=1.5$,
and $\overline{B_z}=0$.
The right eigenvectors for this state are given in Appendix A of \cite{gardiner_unsplit_2008}.

The wave propagates for one period, after which the error is computed between the initial and final state. First we compute the L1 norm, which is the absolute difference for each conserved variable between the initial and final state:

\begin{equation}
    \delta q_s = \frac{1}{n_x n_y n_z} \sum_{i,j,k} \mid q^f_{i,j,k,s} - q^i_{i,j,k,s} \mid,
\end{equation}

\noindent where $q_s$ is a specific conserved variable. We then compute the L2 norm of this vector of L1 norms as

\begin{equation}
    \mid \mid \delta q \mid \mid = \sqrt{\sum_s \left( \delta q_s \right)^2}.
\end{equation}

\begin{figure}[ht!]
    \includegraphics[width=\linewidth]{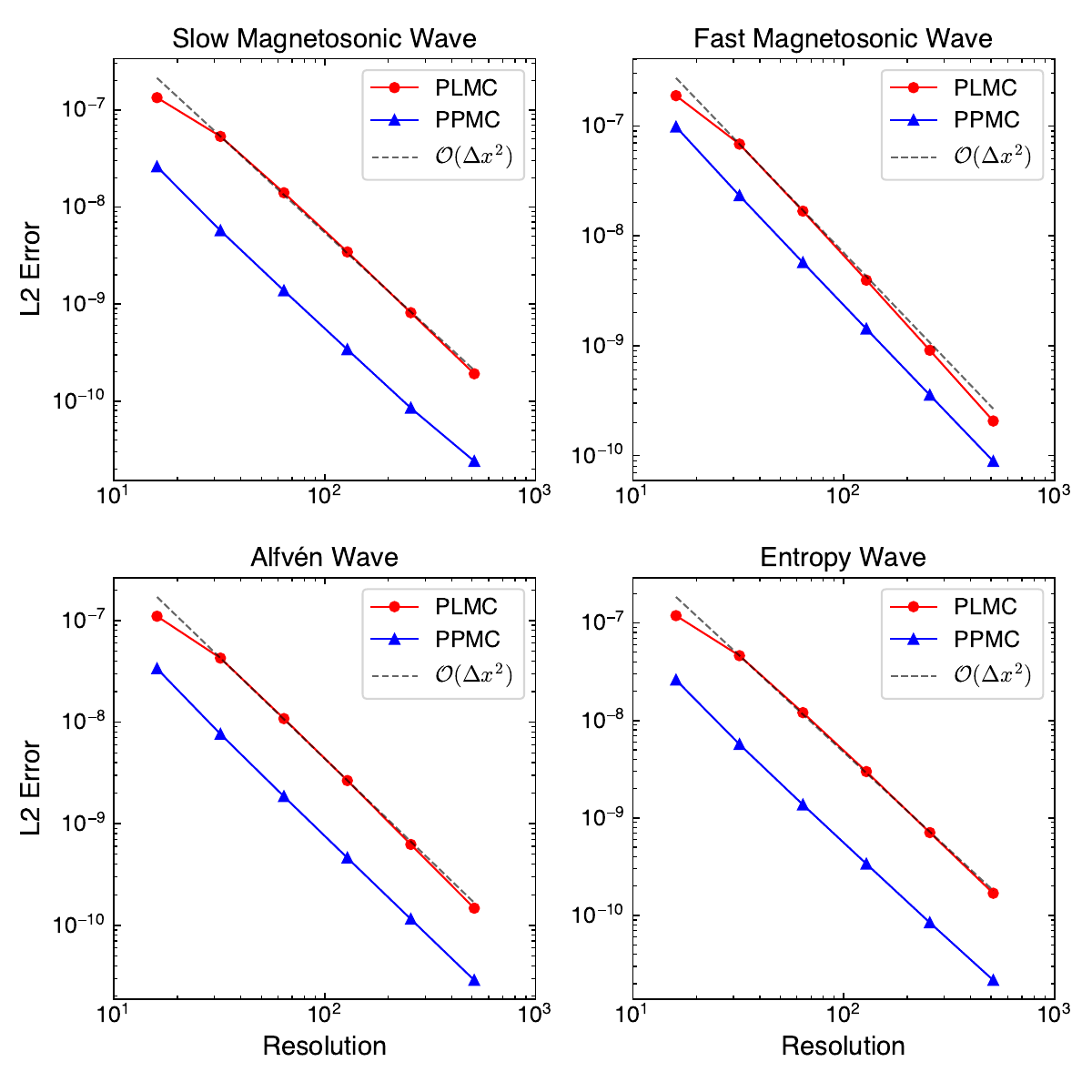}
    \caption{Linear Wave Convergence of all four MHD waves using PLM and PPM reconstruction. \href{https://zenodo.org/records/10927223}{\img{zenodo-gradient-200.png}}}
    \label{fig:linear-wave-convergence}
\end{figure}

These L2 errors are plotted in \autoref{fig:linear-wave-convergence} for both the PLM and PPM reconstructions. The results are comparable to the results in \cite{stone_2009} and demonstrate the expected second order convergence. Using PPM improves the accuracy of the solution by approximately an order of magnitude at any given resolution, but maintains the second order convergence due to the second order nature of the integrator. We have implemented these tests in all three directions with the waves moving in the positive or negative directions and find identical results.

\subsubsection{Circularly Polarized Alfv\'en Wave}
\label{sec:cpaw}

The circularly polarized Alfv\'en wave is a non-linear wave that tests a code's accuracy in the non-linear regime with the quantitative benefits of a regular wave test \citep{Toth1996}. The tests use a periodic domain of $3.0\times1.5\times1.5$ and a resolution of $2N\times N \times N$ where $N$ goes from 8 to 256 in powers of 2. The wave is initialized at an oblique angle the grid, making this a fully 3D test.

\begin{figure}[ht!]
    \includegraphics[width=\linewidth]{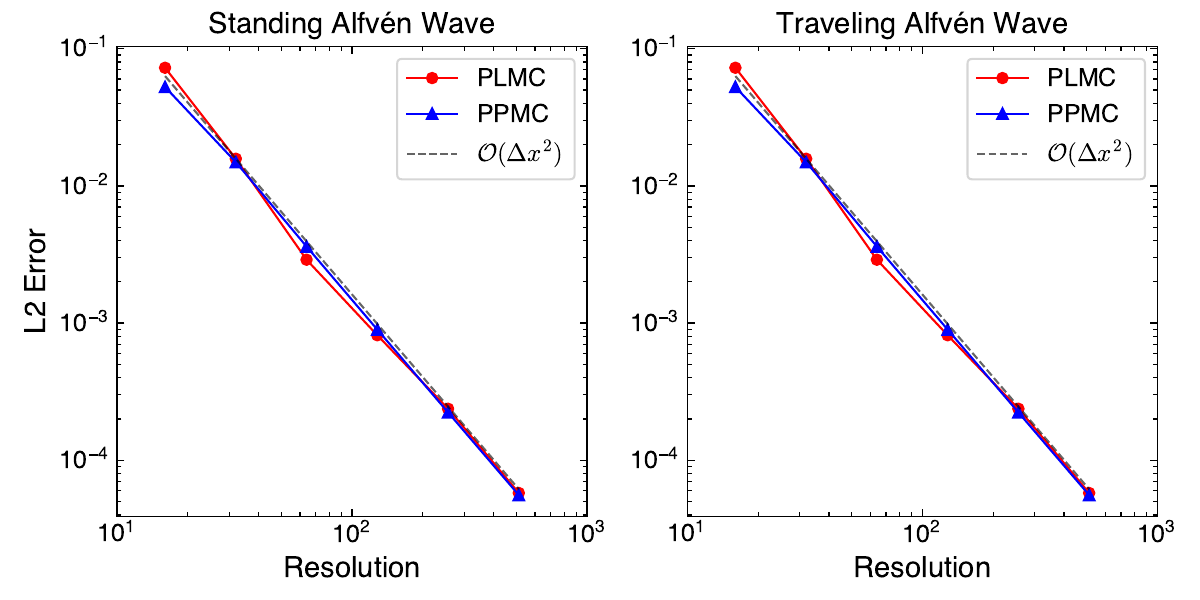}
    \caption{Circularly Polarized Alfv\'en Wave Convergence using PLM and PPM reconstruction. \href{https://zenodo.org/records/10927223}{\img{zenodo-gradient-200.png}}}
    \label{fig:cpaw}
\end{figure}

In a coordinate system aligned with the movement of the wave, the initial conditions are
$\rho = 1.0$,
$P = 0.1$,
$v_x = (0,-1)$ for traveling or standing waves respectively,
$v_y = A \sin{\frac{2\pi x}{\lambda}}$,
$v_z = A \cos{\frac{2\pi x}{\lambda}}$,
$B_x = 1.0$,
$B_y = A \sin{\frac{2\pi x}{\lambda}}$,
and $B_z = A \cos{\frac{2\pi x}{\lambda}}$,
where the amplitude of the wave $A = 0.1$ and the wavelength $\lambda = 1.0$. These coordinates are then transformed with the rotation

\begin{eqnarray}
    x\prime = x \cos\alpha\cos\beta - y \sin\beta - z \sin\alpha\cos\beta \nonumber \\
    y\prime = x \cos\alpha\sin\beta + y \cos\beta - z \sin\alpha\sin\beta \nonumber \\
    z\prime = x \sin\alpha + z \cos\alpha \nonumber
\end{eqnarray}

\noindent with $\sin\alpha = 2/3$ and $\sin\beta = 1/\sqrt{5}$. This ensures the domain is fully periodic through the boundaries and the wave can travel (or stand) indefinitely. The magnetic fields are initialized with the vector potential to ensure initial divergence is zero to round off. The waves are then run for a single period and the L2 norm of the L1 error vector is plotted in Figure \ref{fig:cpaw} using the same method as in Section \ref{sec:lwc}. It is interesting to note that the accuracy improvement of PPM versus PLM seen for the linear waves is absent in this non-linear test.

These Alfv\'en waves are subject to a parametric instability \citep{del_zanna_parametric_2001}, which should not be present for these initial conditions. However, the truncation error will result in small variations in the magnetic pressure which drives low amplitude compression waves \citep{stone_athena_2008}.

\subsubsection{Advecting Field Loop}
\label{sec:afl}

The advecting field loop test consists of a tilted spherical current loop which travels across the domain at an oblique angle to the grid. This test requires particularly accurate balancing of the non-zero components of the induction equation. It also has zero magnetic field outside the spherical current loop; as the current loops moves across the grid those cells that are no longer in the loop should return to zero to within round off error. It is also a good test of the dissipation of the magnetic field, as the magnetic pressure should remain constant.

The initial conditions for this test are most easily described using the magnetic vector potential, which is the vector quantity whose curl equals the magnetic field, $\boldsymbol{B} = \nabla \times\boldsymbol{A}$. The background state is
$\rho = 1.0$,
$P = 1.0$,
$v_x = 1.0$,
$v_y = 1.0$,
$v_z = 2.0$,
$B_x = 0$,
$B_y = 0$, and
$B_z = 0$.

In the central region the state is given by the following vector potential, which we have chosen such that $A_x = 0$:
\begin{equation}
    A_y = A_z =
    \begin{cases}
        A \left( R - r \right),& \text{for}\; r < R\\
        0,              & \text{otherwise}
    \end{cases}
\end{equation}

\noindent where $r$ is the Euclidean distance from the center of the domain, $R = 0.3$, and the amplitude $A=10^{-3}$. Note that since the vector potential is along the vertices of the cells, $A_y$ and $A_z$ will never have the same value at the same position as they are not stored at identical positions. The test is conducted on a grid of $N\times N\times 2N$ cells for $N=(32, 64, 128, 256)$ with a periodic domain of $1.0\times1.0\times2.0$ centered at zero and evolved for two periods; $t_{max} = 2.0$.

\begin{figure}[ht!]
    \includegraphics[width=\linewidth]{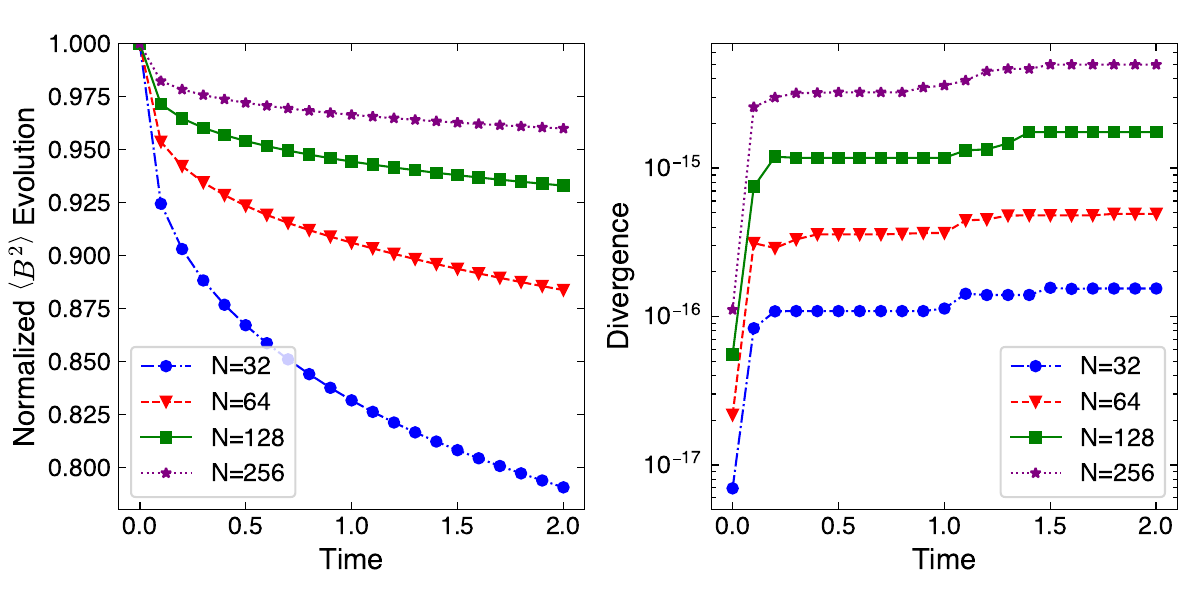}
    \caption{Evolution of tilted spherical magnetic field loop through two full periods using PPMC reconstruction}. Mean of $B^2$ normalized to the initial value as a function of time (left) and the maximum divergence in the domain as a function of time (right). \href{https://zenodo.org/records/10927223}{\img{zenodo-gradient-200.png}}
    \label{fig:afl}
\end{figure}

\begin{figure}[ht!]
    \includegraphics[width=\linewidth]{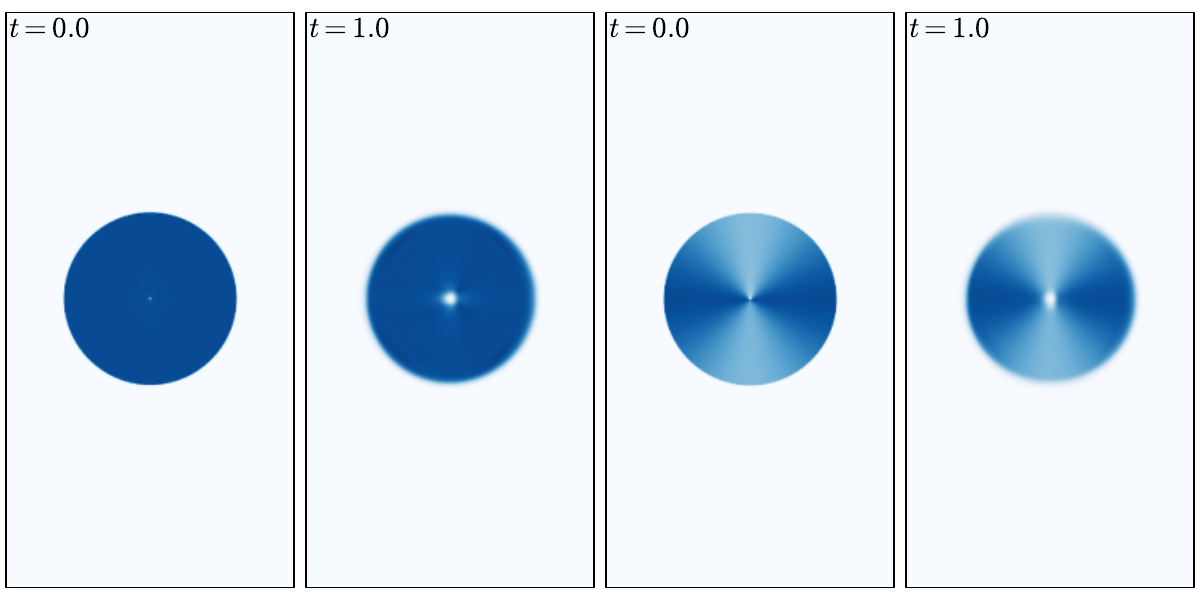}
    \caption{Cross sections of the spherical advecting field loop magnetic energy density at $t=0.0$ and one period. The first and second panels show a slice centered on the loop through the plane of symmetry. The third and fourth panels show a slice along the $x-z$ plane. Note that these figures utilize PLMC reconstruction as PPMC introduced spurious oscilations in the direction of advection. \href{https://zenodo.org/records/10927223}{\img{zenodo-gradient-200.png}}}
    \label{fig:afl_slice}
\end{figure}

Figure \ref{fig:afl} shows the mean of cell centered $B^2$, normalized to the initial value, in order to demonstrate the convergence of the dissipation rate. The dissipation rate is comparable to those found in the literature \citep{stone_athena_2008} and improves at approximately first order. Figure \ref{fig:afl} also shows the maximum divergence in the domain as a function of time. Throughout the entire evolution it remains near round off and, after an initial rise, remains fairly constant. The zero magnetic field region outside of the current loop also remains near zero throughout the entire evolution of the problem. Figure \ref{fig:afl_slice} shows cross sections of the loop initial conditions and after one period with a resolution of $128\times128\times256$ cells. The shape is well maintained with minimal dissipation.

\subsubsection{MHD Riemann Problems}
\label{sec:riemann}

MHD Riemann problems are standard tests for new MHD codes and methods due to their mix of different flow types and extreme conditions. There are many different MHD Riemann problems in the literature \citep{brio_wu_1988, einfeldt_1991, ryu_jones_1995, dai_woodward_1998}; we have chosen to present five that cover a variety of challenging cases.

The typical Riemann problem setup uses a domain with a discontinuity at the midpoint between two different states. As the problem evolves in time, the waves propagate along characteristics such that the solution evolves self-similarly in time. All of Riemann problems shown in this section employ a domain of $1\times1\times1$ and resolution of $512\times16\times16$ and are run until the $t_{max}$ which is specified for that particular problem. We use parabolic reconstruction with limiting in the characteristic variables unless otherwise noted. All have been run in all three spatial directions with both possible orientations of the two states and achieved identical results. The details of each left and right state are given in Table \ref{table:riemann}.


\begin{table}
    \centering
    \begin{tabular}{lcccccccccc}




        Riemann Problem & $\gamma$ & $t_{max}$ & $B_x$ & $\rho_L$ & $P_L$ & $v_{x,L}$ & $v_{y,L}$ & $v_{z,L}$ & $B_{y,L}$ & $B_{z,L}$  \\ \hline

        Brio \& Wu           & 2             & 0.1  & 0.75                    & 1    & 1    & 0    & 0    & 0   & 1                         & 0                       \\ \hline
        Dai \& Woodward      & $\frac{5}{3}$ & 0.2  & $\frac{2}{\sqrt{4\pi}}$ & 1.08 & 0.95 & 1.2  & 0.01 & 0.5 & $\frac{3.6}{\sqrt{4\pi}}$ & $\frac{2}{\sqrt{4\pi}}$ \\ \hline
        Ryu \& Jones 1a      & $\frac{5}{3}$ & 0.08 & $\frac{5}{\sqrt{4\pi}}$ & 1    & 20   & 10   & 0    & 0   & $\frac{5}{\sqrt{4\pi}}$   & 0                       \\ \hline
        Ryu \& Jones 4d      & $\frac{5}{3}$ & 0.16 & 0.7                     & 1    & 1    & 0    & 0    & 0   & 0                         & 0                       \\ \hline
        Einfeldt Rarefaction & 1.4           & 0.16 & 0                       & 1    & 0.45 & -2   & 0    & 0   & 0.5                       & 0                       \\ \hline



    \end{tabular}
    \newline
    \newline
    \newline
    \begin{tabular}{lcccccccc}




        Riemann Problem & $\rho_R$ & $P_R$ & $v_{x,R}$ & $v_{y,R}$ & $v_{z,R}$ & $B_{y,R}$ & $B_{z,R}$ \\ \hline

        Brio \& Wu           & 0.125 & 0.1  & 0     & 0   & 0   & -1                      & 0                       \\ \hline
        Dai \& Woodward      & 1     & 1    & 0     & 0   & 0   & $\frac{4}{\sqrt{4\pi}}$ & $\frac{2}{\sqrt{4\pi}}$ \\ \hline
        Ryu \& Jones 1a      & 1     & 1    & -10   & 0   & 0   & $\frac{5}{\sqrt{4\pi}}$ & 0                       \\ \hline
        Ryu \& Jones 4d      & 0.3   & 0.2  & 0     & 0   & 1   & 1                       & 0                       \\ \hline
        Einfeldt Rarefaction & 1     & 0.45 & 2     & 0   & 0   & 0.5                     & 0                       \\ \hline



    \end{tabular}
    \caption{Riemann Problem Initial Conditions.} The $L/R$ subscripts indicate that it is the left/right state. $B_x$ is always the same in both states.
    \label{table:riemann}
\end{table}

\begin{figure}[ht!]
    \includegraphics[width=\linewidth]{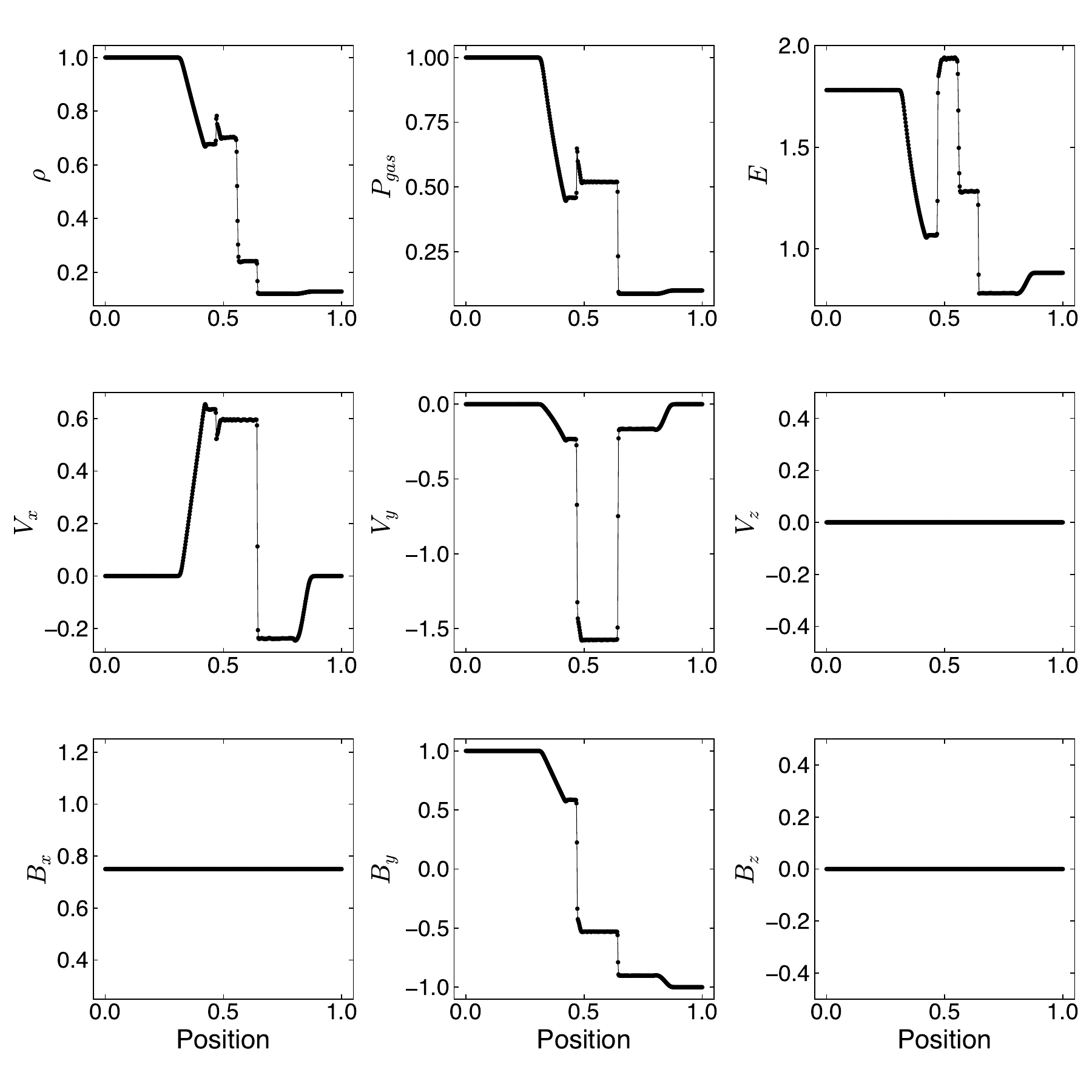}
    \caption{The Brio \& Wu Shock Tube solution \citep{brio_wu_1988}.
    \href{https://zenodo.org/records/10927223}{\img{zenodo-gradient-200.png}}}
    \label{fig:brio-and-wu}
\end{figure}

\paragraph{Brio \& Wu Shock Tube}
Figure \ref{fig:brio-and-wu} shows the Brio \& Wu Shock Tube \citep{brio_wu_1988} which is a staple test of MHD codes. This Riemann problem is essentially the Sod shock tube \citep{sod_1978} with a magnetic field. However, this shock tube is an excellent stress test for PPM reconstruction, as methods higher than second order tend to create large oscillations in the solution due to the slowly moving shock waves. We have implemented both the PPM reconstruction algorithm used in the VL+CT integrator from the method described in \cite{stone_athena_2008} as well as the method described in \cite{felker_2018}, and find that the latter significantly reduces oscillations for this test\textbf{; Figure \ref{fig:brio-and-wu} uses the method from \cite{felker_2018}, which is the version currently available in Cholla}. With PPM reconstruction, we find  oscillations in the solution when limiting in either the primitive or characteristic variables. No oscillations are present when using PLM reconstruction.

\begin{figure}[ht!]
    \includegraphics[width=\linewidth]{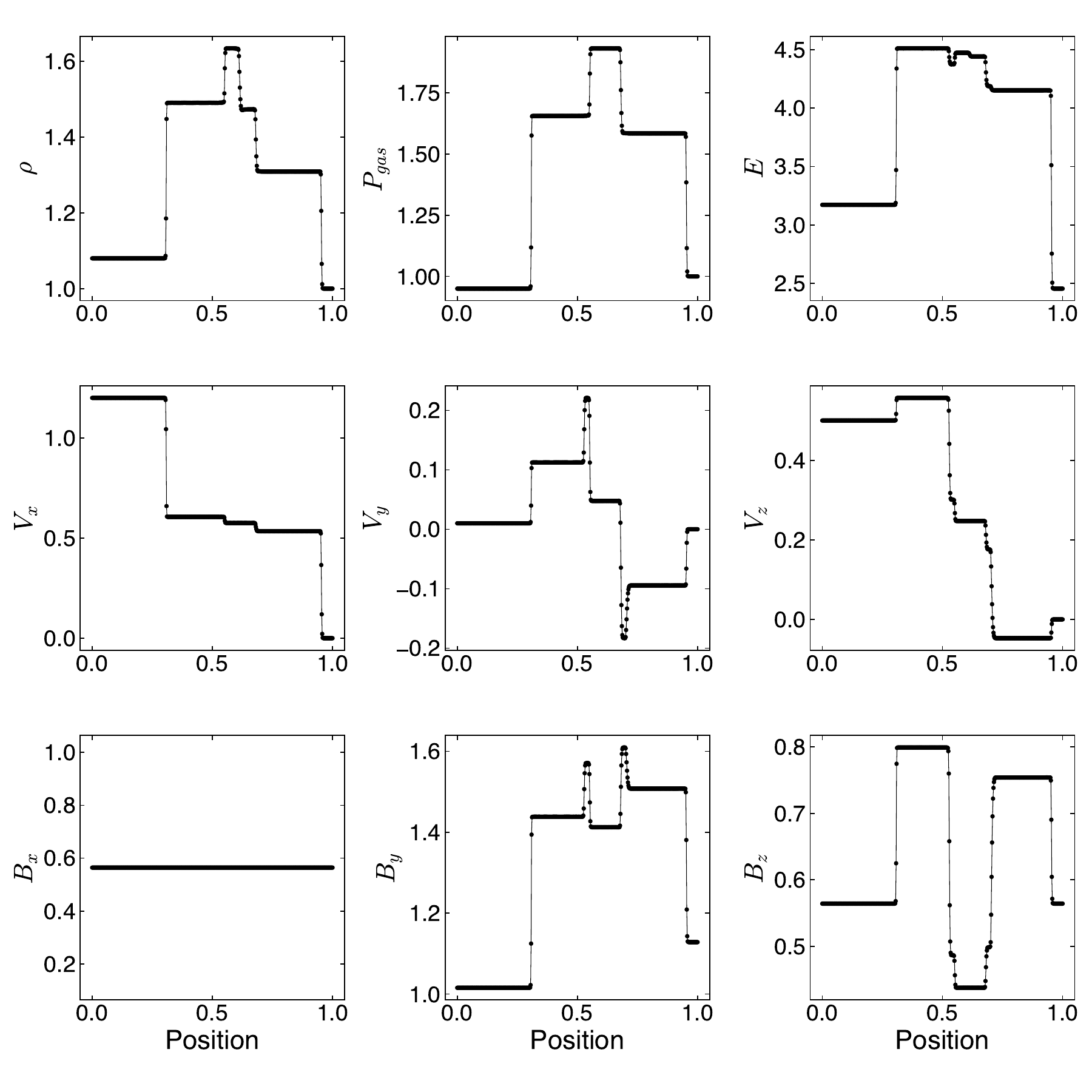}
    \caption{Dai \& Woodward Shock Tube (also called Ryu \& Jones 2a) solution \citep{dai_woodward_1998, ryu_jones_1995}.
    \href{https://zenodo.org/records/10927223}{\img{zenodo-gradient-200.png}}}
    \label{fig:dai-and-woodward}
\end{figure}

\paragraph{Dai \& Woodward Shock Tube}
Figure \ref{fig:dai-and-woodward} shows the Dai \& Woodward Shock Tube (also called Ryu \& Jones 2a) \citep{dai_woodward_1998, ryu_jones_1995} which produces all seven possible MHD waves. From left to right they are: fast shock, Alfvén wave, slow shock, contact discontinuity, slow shock, Alfvén wave, and fast shock. This makes it an excellent laboratory for checking that the full spread of wave modes are well resolved.

\begin{figure}[ht!]
    \includegraphics[width=\linewidth]{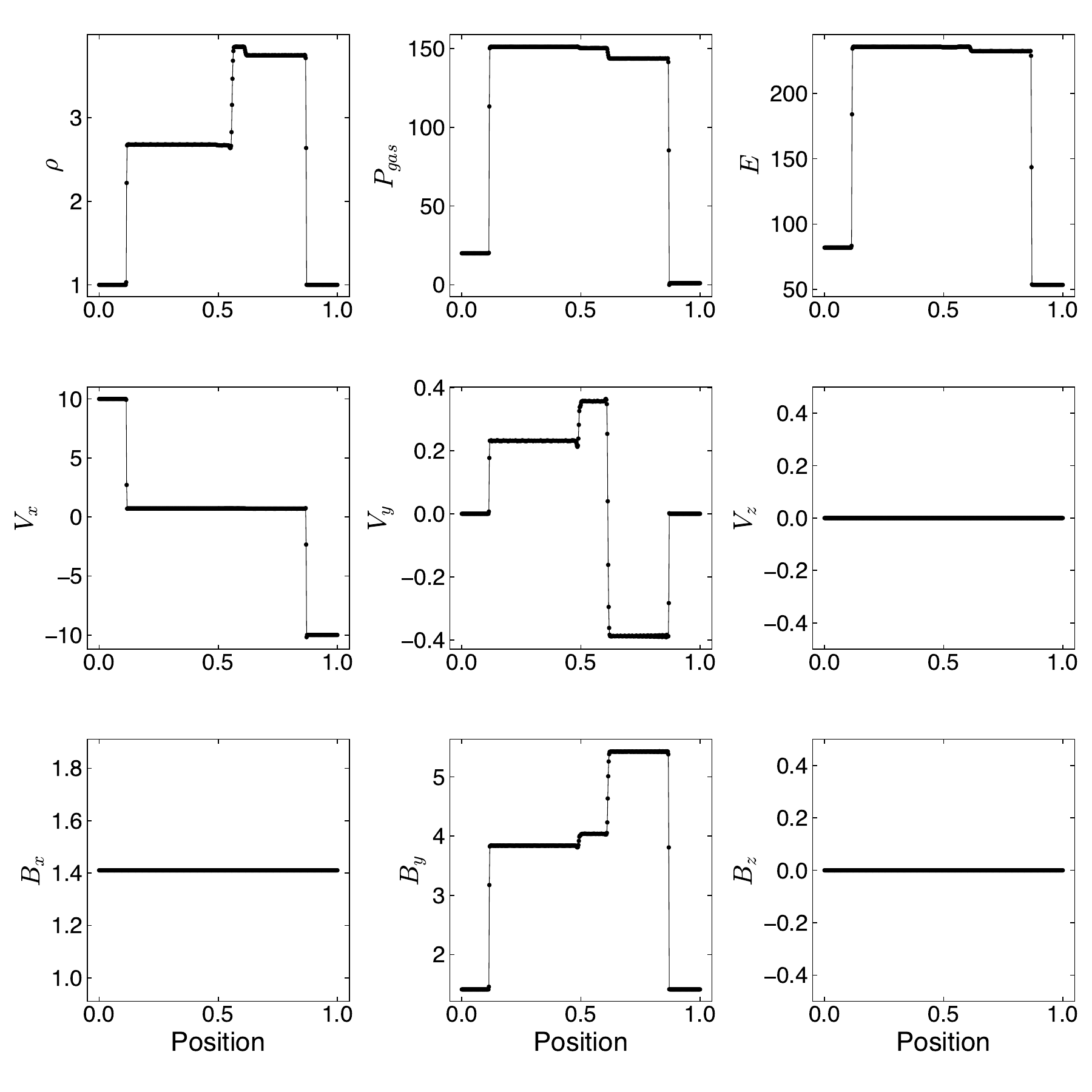}
    \caption{Ryu \& Jones 1a Shock Tube solution \citep{ryu_jones_1995}.
    \href{https://zenodo.org/records/10927223}{\img{zenodo-gradient-200.png}}}
    \label{fig:rj-1a}
\end{figure}

\paragraph{Ryu \& Jones 1a Shock Tube}
Figure \ref{fig:rj-1a} shows the Ryu \& Jones 1a Shock Tube solution \citep{ryu_jones_1995} which is a less common test for MHD codes. However, in our experience, it is an excellent problem for debugging due to its relatively simple structure, which is easy to examine manually. The lack of any spikes and the presence of multiple types of strong shocks also make it a good diagnostic test for over/undershoot of the solution near discontinuities.

\begin{figure}[ht!]
    \includegraphics[width=\linewidth]{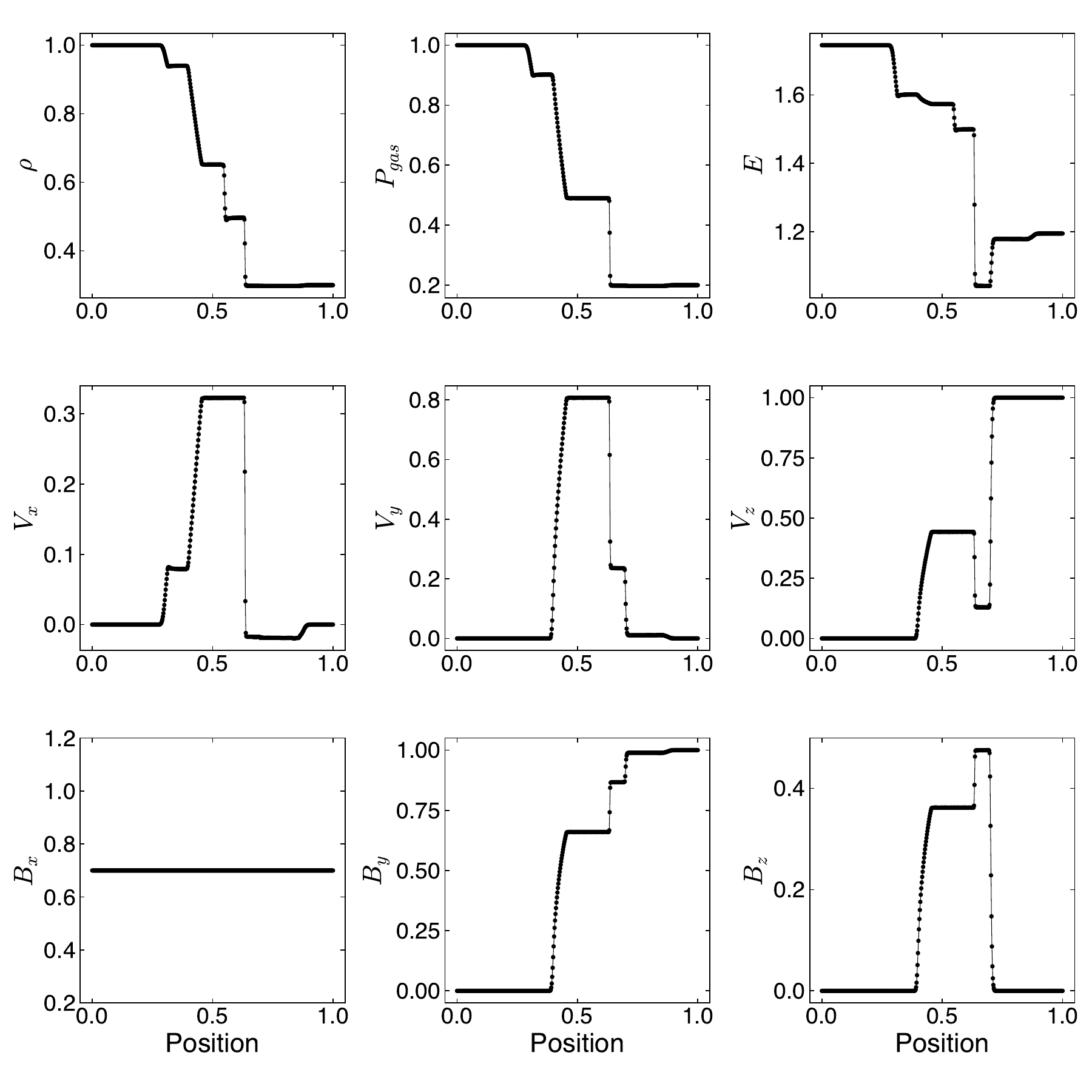}
    \caption{Ryu \& Jones 4d Shock Tube solution \citep{ryu_jones_1995}.
    \href{https://zenodo.org/records/10927223}{\img{zenodo-gradient-200.png}}}
    \label{fig:rj-4d}
\end{figure}

\paragraph{Ryu \& Jones 4d Shock Tube}
Figure \ref{fig:rj-4d} shows the Ryu \& Jones 4d Shock Tube solution \citep{ryu_jones_1995} which features a switch-on slow shock. Switch-on waves increase the strength of the transverse magnetic field while reducing the thermal pressure to maintain energy conservation. This is a simplified example of a type of magnetic field amplification and as such it is important to demonstrate that a code can replicate it accurately. A switch-off wave does the inverse.

\begin{figure}[ht!]
    \includegraphics[width=\linewidth]{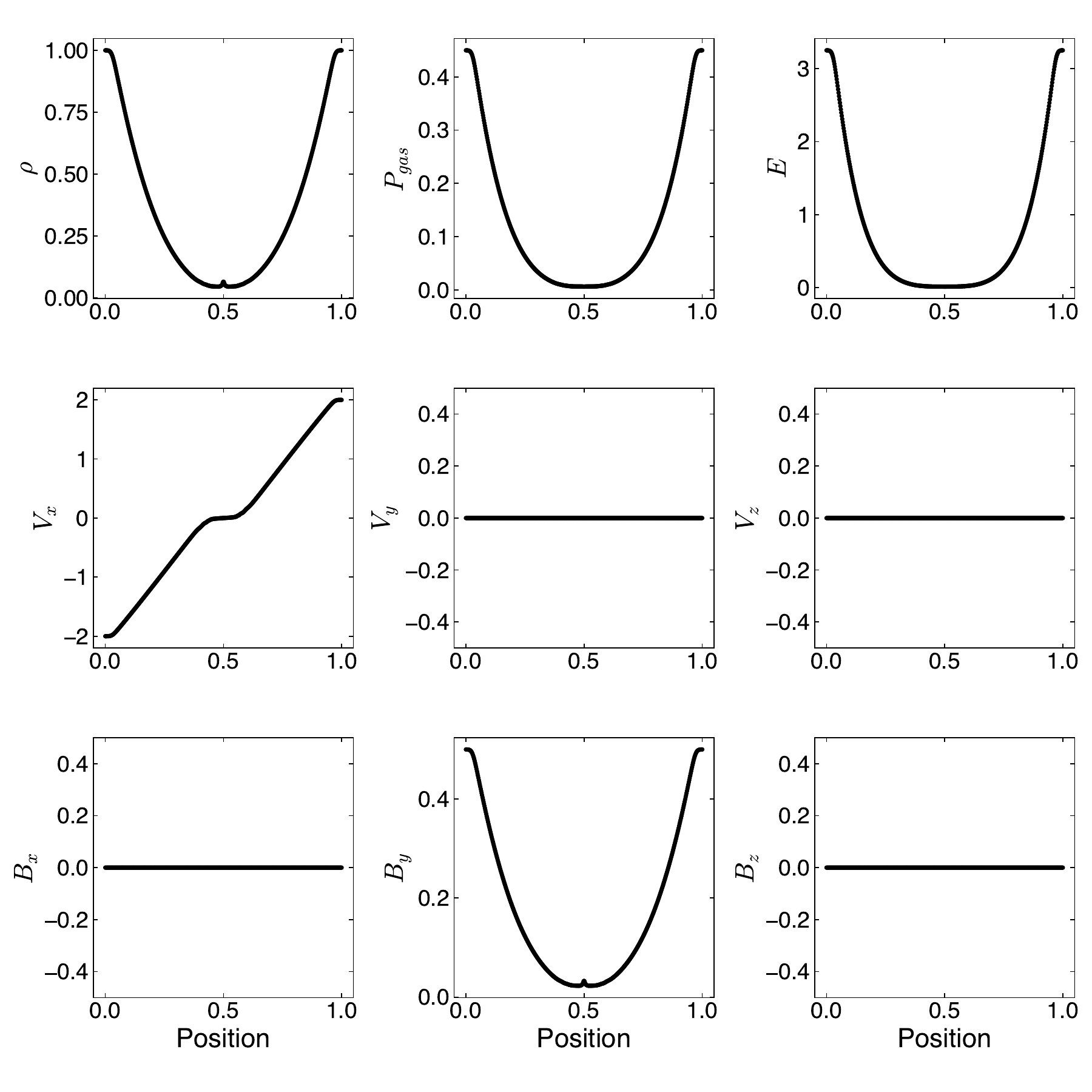}
    \caption{MHD Einfeldt Strong Rarefaction solution \citep{einfeldt_1991}.
    \href{https://zenodo.org/records/10927223}{\img{zenodo-gradient-200.png}}}
    \label{fig:einfeldt}
\end{figure}

\paragraph{MHD Einfeldt Strong Rarefaction}
Figure \ref{fig:einfeldt} shows the MHD Einfeldt Strong Rarefaction test \citep{einfeldt_1991} which creates a strong outflow and central vacuum state. The diverging solution leads to an extremely strong and fast rarefaction where the energy is dominated by kinetic energy and as such can often reveal challenges for finite-volume methods with near-vacuum states, since some Riemann solvers will return unphysical solutions with negative density or negative internal energy. High values of the outflow velocity ($V_{out}\ge3$) can also lead to spurious oscillations in the solution. $V_{out} = 2$ was chosen for this test \citep{charm_2011}. Cholla performs well on this test with no spurious oscillations or unphysical negative values.

\subsubsection{MHD Blast Wave in a Strongly Magnetized Medium}
\label{sec:mhd-blast}

Blast waves in different forms are excellent tests for hydrodynamics and MHD codes. They combine strong shocked flows, smooth flows, and, in MHD, strong magnetic fields. The results are qualitative rather than quantitative, but thoroughly test the robustness of the algorithm and are excellent regression tests for automated testing (see Section \ref{sec:testing}). For this test we use $\beta = 0.2$; like \cite{stone_2009}, we find instabilities if $\beta$ is decreased by a factor of 10.

\begin{figure}[ht!]
    \includegraphics[width=\linewidth]{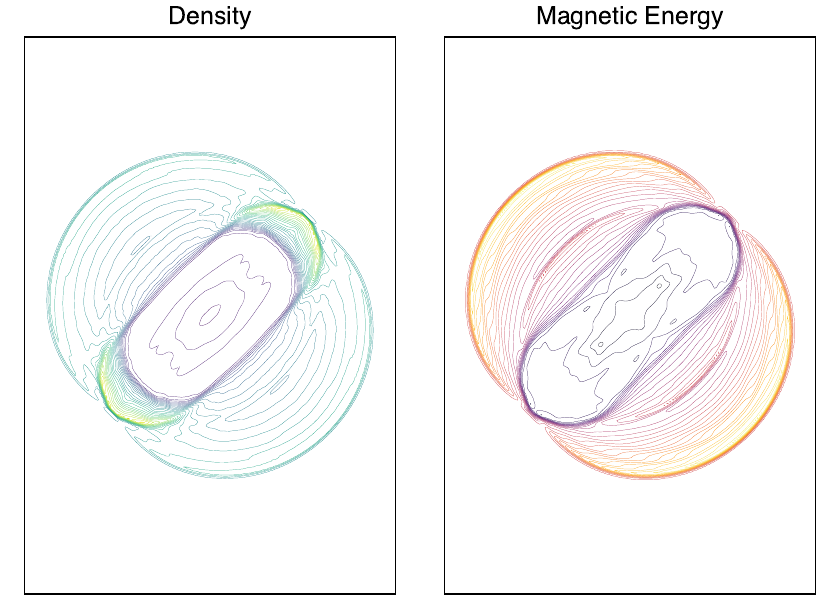}
    \caption{Contour plot of the MHD blast wave test at $t=0.2$. 30 evenly spaced contours are shown in an $x-y$ slice through the center of the domain. \href{https://zenodo.org/records/10927223}{\img{zenodo-gradient-200.png}}}
    \label{fig:blast}
\end{figure}

The background state is
$\rho = 1.0$,
$P = 0.1$,
$v_x = 0.0$,
$v_y = 0.0$,
$v_z = 0.0$,
$B_x = 1/\sqrt{2}$,
$B_y = 1/\sqrt{2}$,
$B_z = 0.0$,
and the over pressure region is a central sphere of size $R = 0.1$ which has $P=10.0$. The test is then run on a domain of $1\times1.5\times1$ with a resolution of $200\times300\times200$ cells until $t = 0.2$. Figure \ref{fig:blast} shows contours of the density and magnetic energy fields in an $x-y$ slice through the center of the domain. The contours are smooth and symmetric and show clear elongation of the blast wave rarefaction parallel to the magnetic field. The blast wave propagates slowly parallel to the magnetic field but much more rapidly perpendicular to the magnetic field.

\subsubsection{Orszag-Tang Vortex}
\label{sec:otv}

The Orszag-Tang vortex is a standard 2D MHD test from \cite{otv_1979}. While it does not provide a quantitative measure of accuracy like the linear wave tests or a test of the robustness of the method like the MHD blast wave, it does have a very complex flow that is sensitive to changes in the integrator, making it ideal for regression testing.

\begin{figure}[!ht]
    \includegraphics[width=\linewidth]{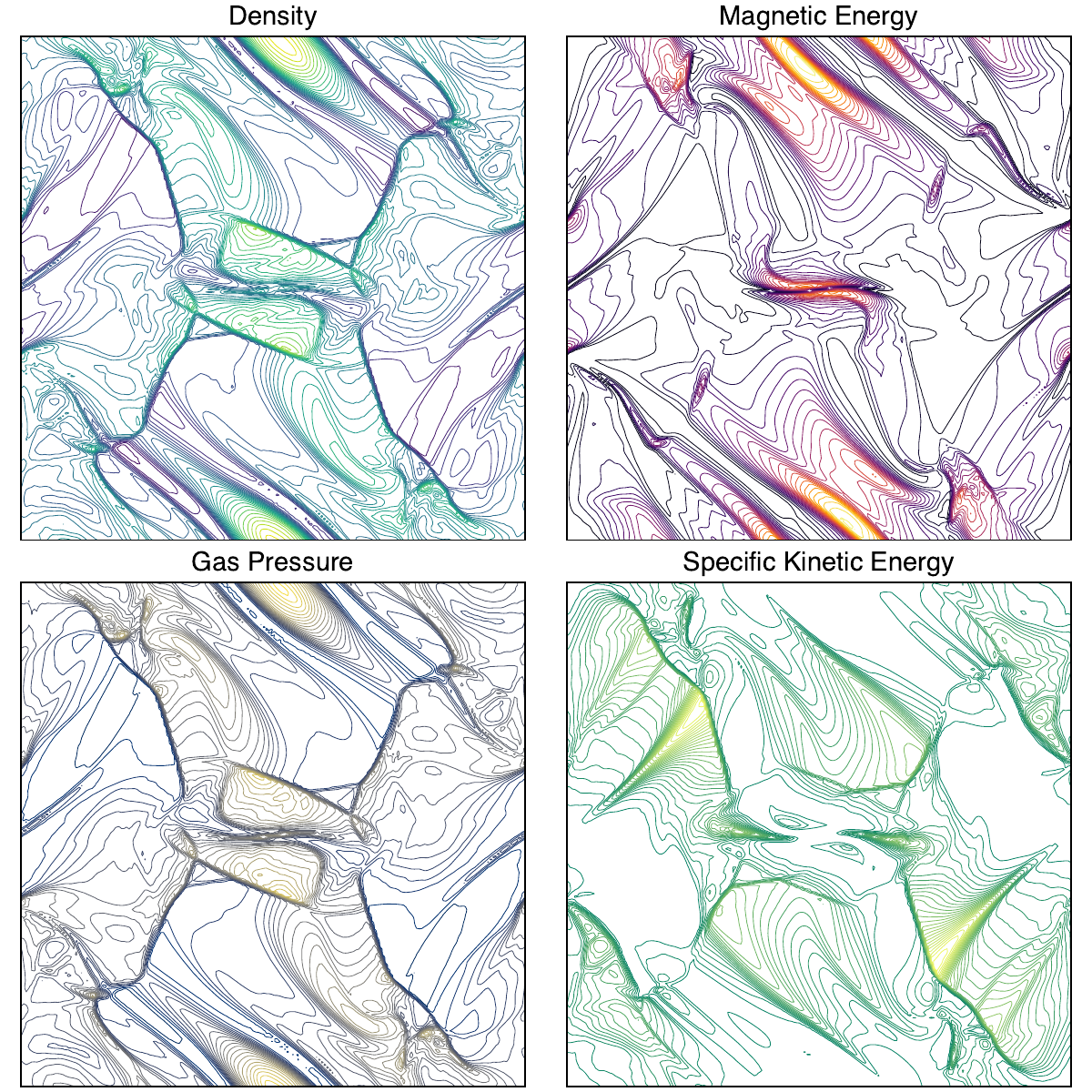}
    \caption{Contour plot of the Orszag-Tang Vortex at $t=0.5$. Thirty evenly spaced contours are shown for each plot in an $x-y$ slice through the center of the domain.  \href{https://zenodo.org/records/10927223}{\img{zenodo-gradient-200.png}}}
    \label{fig:otv}
\end{figure}

The test was conducted on a periodic domain of $1\times1\times1$ with a resolution of $192\times192\times192$ cells until $t = 0.5$ with the following initial conditions:
$\rho = 25 / \left( 36 \pi \right)$,
$P    =  5 / \left( 12 \pi \right)$,
$v_x  = \sin 2\pi y$,
$v_y  = -\sin 2\pi x$,
$v_z  = 0.0$,
$A_x  = 0.0$,
$A_y  = 0.0$,
$A_z  = \left( B_0/4\pi \right) \left( \cos{4\pi x} + 2 \cos{2\pi y} \right)$, with $B_0 = 1\sqrt{4\pi}$.
The results, plotted in Figure \ref{fig:otv}, can be compared directly to Figure 22 in \cite{stone_athena_2008} as a qualitative check for correctness of the flow structure.

\subsection{MHD Performance Tests}
\label{sec:mhd-perf-tests}

\begin{figure}[ht!]
    \epsscale{0.5}
    \includegraphics[width=0.5\linewidth]{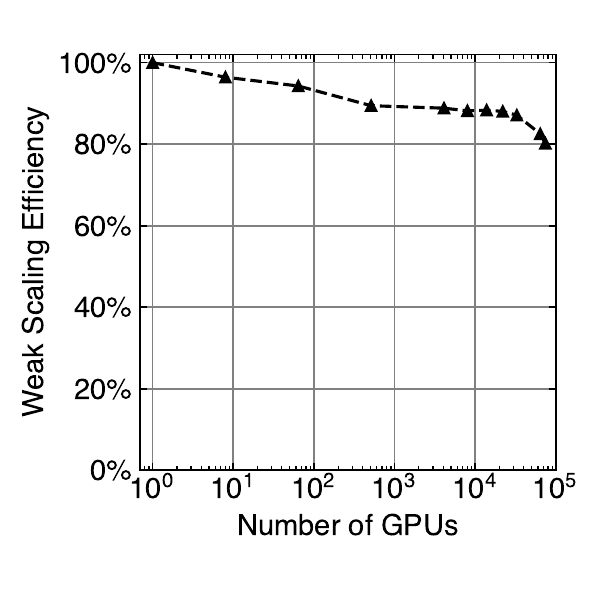}
    \includegraphics[width=0.5\linewidth]{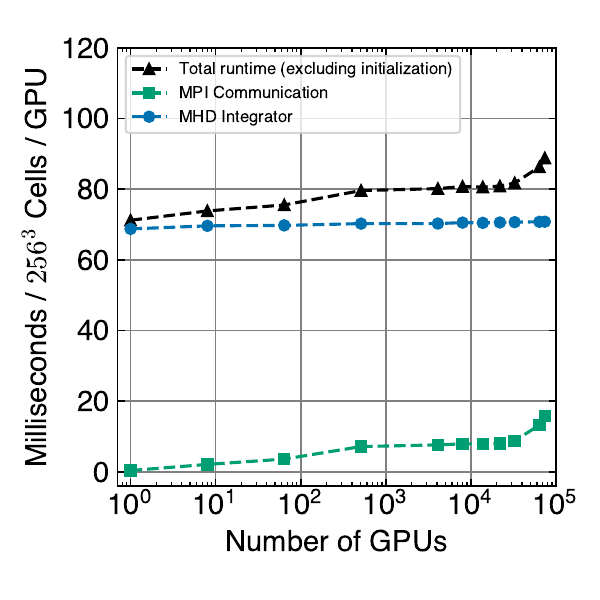}
    \caption{Weak scaling performance of Cholla MHD. When running on a single GPU Cholla updates $2.04\times10^8$ cells per second per GPU; the largest run with 74,088 GPUs updates $1.67\times10^8$ cells per second per GPU, a weak scaling efficiency of 82.2\%. The 74,088 GPU run updates a total of $1.24\times10^{13}$ cells per second. \href{https://zenodo.org/records/10927223}{\img{zenodo-gradient-200.png}}}
    \label{fig:scaling-weak-efficiency}
\end{figure}

\begin{figure}[ht!]
    \centering
    \includegraphics{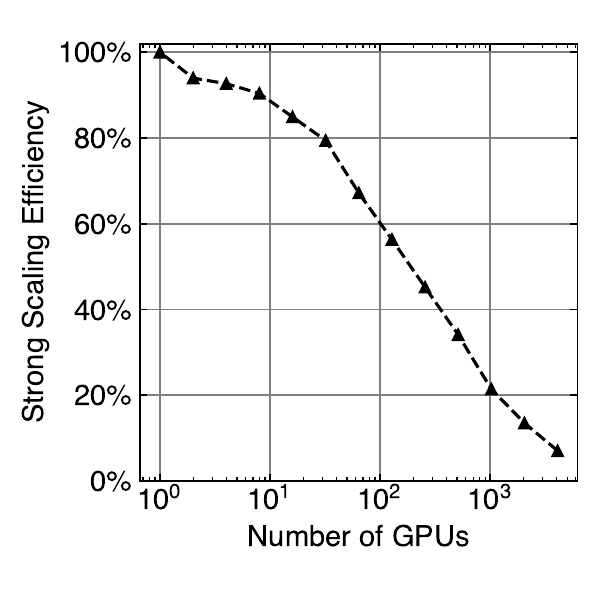}
    \caption{Strong scaling performance of Cholla MHD with a problem size of $459^3$ cells. \textbf{Note that a typical number of cells per GPU is $>128^3$ which is corresponds to 64 GPUs in this test.} \href{https://zenodo.org/records/10927223}{\img{zenodo-gradient-200.png}}}
    \label{fig:scaling-strong}
\end{figure}

Given that Cholla is a massively parallel code, its scaling properties warrant discussion. Our primary focus is on weak scaling rather than strong scaling, since good weak scaling enables much larger problems to be simulated, while strong scaling can lead to reducing the number of cells per GPU to the point where the whole GPU cannot be utilized, which will significantly impact performance. Results of our weak scaling tests are shown in Figure \ref{fig:scaling-weak-efficiency}, while strong scaling is shown in Figure \ref{fig:scaling-strong}.

All of the weak scaling tests shown were performed with a slow magnetosonic wave perturbation (described in Section \ref{sec:lwc}), periodic boundary conditions, in double precision, and with $459^3$ cells per MPI rank; each rank is assigned one GPU. We employ the second order piecewise linear reconstruction method with limiting in the characteristic variables. The wave is evolved for 100 time steps (a wall-clock time of $\sim45$ seconds) and the resulting time per step is averaged over the total number of time steps, excluding setup and tear down time. Strong scaling tests are performed with the same problem, though the wave is run through one full period, $\sim 6100$ time steps to ensure that it runs for a non negligible wall-clock time when large numbers of GPUs are utilized. The strong scaling test uses $459^3$ cells, the most we can fit on a single MI250X GCD.

Our scaling tests were performed on the \textit{Frontier} Supercomputer at the Oak Ridge Leadership Computing Facility. \textit{Frontier} utilizes AMD MI250X GPUs, each of which contains two Graphics Compute Dies (GCDs) that largely function as separate GPUs and can be treated as such in software. Thus, for the sake of clear comparison to other systems, we will refer to each GCD as a single GPU for the remainder of this paper.

On \textit{Frontier} Cholla updates $2.36\times10^8$ cells per second per GPU when running with a single GPU. The single-GPU performance is comparable on NVIDIA hardware: an NVIDIA V100 GPU achieved 160 million cell updates per second, and an A100 achieved 259 million cell updates per second. On 74,088 GPUs, nearly the entirety of \textit{Frontier}, Cholla performs $1.89\times10^8$ cell updates per second per GPU, with a weak scaling efficiency of 82.2\%. The 74,088 GPU run updated a total of $1.40\times10^{13}$ cells per second on a total grid size of $19,278^3$ cells. This performance is comparable to other similar optimized GPU codes (see e.g. Figures 9 and 10 in \cite{parthenon_2023}).\footnote{Note that the weak scaling plots in \cite{parthenon_2023} are normalized to single node performance, not single GPU performance, and they report performance for a 2nd order hydrodynamic solver, not MHD.}

With a single MPI rank there is negligible communication overhead as no halo cells need to be exchanged between GPUs; the only boundary update required is to copy halo cell values from one location in GPU memory to another. This update is very fast, and accounts for $\sim 0.6\%$ of total runtime on a single rank. As the number of ranks grows the MPI overhead quickly stabilizes at around 15ms with a moderate increase when running on the full size of \textit{Frontier}. Perhaps most importantly, the VL+CT integrator scales almost perfectly and takes up most of the time on each time step, dominating over the MPI communication time by nearly a factor of 10. The very slightly imperfect weak scaling of the integrator, which has no MPI communication, is due to GPU-to-GPU variance; as the number of GPUs increases the standard deviation in runtime for a single time step increases. Since all processes must wait for the slowest one to complete before beginning the next boundary exchange, this naturally leads to a few percent decrease in efficiency on these scales. Overall these tests demonstrate that Cholla has excellent weak scaling up to the full size of \textit{Frontier}, over 74,000 GPUs.

\textbf{Strong scaling is also an important metric in determining the optimally sized problem per GPU}. Cholla's strong scaling performance (Figure \ref{fig:scaling-strong}) is close to ideal up to 32 GPUs (80\% strong scaling efficiency) and does not drop below 50\% until 256 GPUs. Strong scaling plateaus at around 512 GPUs, where the problem size per GPU is $58^3$ cells. Running with a typical number of cells ($128^3 - 256^3$) to balance wall-time versus efficiency, we suffer at most a $\approx55\%$ drop in efficiency. (In this strong scaling plot the closest problem size to $128^3$ is the 64 GPU test, which has $115^3$ cells per GPU and a strong scaling efficiency of 67.2\%)


\section{Automated Testing \& Continuous Integration}
\label{sec:testing}

As Cholla has continued to grow in complexity, and with the continued addition of large new physics modules like MHD that require changes to much of the code base, the need for a more robust, automated testing system has become increasingly apparent. Several challenges exist in implementing such a system for Cholla - not only must the testing system accommodate GPU hardware, but it must also work on large parallel systems like \textit{Frontier}. We have addressed this need in three primary steps: 1) Choosing a testing framework for unit tests, 2) Writing the software tools needed to extend that testing framework for Cholla's needs, and 3) adding automated testing as part of a continuous integration (CI) pipeline that runs whenever anyone submits a pull request to the Cholla GitHub repository. This Section describes the implementation of our automated testing and continuous integration framework.

Taken together, the ability to add tests for any part of the code and their automated running on every pull request has meant that new features are faster and easier to add to Cholla with confidence. Errors or changes that may otherwise break older code are often caught by the tests before they ever make it into the code base.

\subsection{Unit Testing Framework}
\label{sec:testing-framework}

Three main kinds of tests are needed for scientific code bases: unit tests, integration tests, and system tests (also called ``end-to-end" tests). Unit tests check a single ``unit" of code, i.e. a single function, class method, data structure, etc. Integration tests check how units of code operate together; for example, a test of the HLLD solver is an integration test because the solver internally calls many different functions and uses multiple different data structures. System tests check the entire code base, often with a simple test problem like a Riemann problem or linear wave, and verify that the entire program produces the correct output.

We chose GoogleTest\footnote{\url{https://github.com/google/googletest}} for the unit testing framework due to its large number of features, general popularity, and relative ease of use. Another primary requirement was a testing framework that supports ``death tests", i.e. tests that check if the internals of the test crash/segfault/etc. Since Cholla, like many high performance computing (HPC) codes, handles errors by reporting those errors and then exiting, a testing framework that can handle code crashes without crashing the tests as well is critical. GoogleTest is available already built on many HPC systems and, if it is not available, can be built as an optional part of the test running script.

\subsection{Extensions for Cholla}

Two primary extensions were required for fully testing Cholla: a robust method for comparing floating point numbers and a way to run system tests.

\subsubsection{Floating Point Comparisons}
\label{sec:fp-comparing}

In order to run either unit tests or system tests, the code must have a robust method for comparing floating point numbers for equality. Comparing floating point numbers for equality is notoriously challenging and generally the exact comparison method that should be used varies by application\citep{goldberg_1991,muller_2018}\footnote{\url{https://randomascii.wordpress.com/2012/02/25/comparing-floating-point-numbers-2012-edition/}}. Absolute comparisons ($|a-b| < X$) work well for small numbers, but with larger numbers the difference between two successive floats can be much larger than a typical value for $X$ and therefore a different comparison method is required. We chose a hybrid method of both an absolute comparison and a Units in Last Place (ULP) comparison. A ULP comparison determines how many representable floating point numbers there are between any two floats. By default the hybrid method we employ first performs an absolute check, $|a-b| < 10^{-14}$. This number was chosen as we found that typical differences in the Sod Shock Tube solution when comparing results with different hardware and compilers resulted in differences of $\sim5\times10^{-15}$. After the absolute check a ULP check is performed with a maximum allowed error of 4. If either check passes then the numbers are deemed to be ``equal". Having a robust method to compare floating point numbers is critical since the output of a code is not guaranteed to be bitwise identical when the code is compiled with different compilers, run on different systems, etc.

\subsubsection{System Tests}

GoogleTest provides most of the tools required to run unit and integration tests, but validating the results of an entire system test requires additional infrastructure. Running a system test requires launching the program to test with correct initial conditions, checking that the program did not crash, loading both the generated data to test and the fiducial data, then comparing those two data sets. In order to perform system tests with Cholla, we added a class that performs all of the required tasks, which include launching Cholla with any number of MPI ranks as well as comparing the results against fiducial data. To facilitate running across a wide range of MPI ranks and on clusters with queue systems, the class is designed to allow system tests to be run in different modes: one can either launch Cholla and save the results, compare already existing test data to fiducial data, or do both. This enables the user to run Cholla on many thousands of ranks then later launch a separate job to make the actual comparison. We have found that on up to 10,000 ranks, with small simulation grids (typically $<64^3$ cells) per rank the latter comparison only takes a few minutes on a single CPU core. Most of the time, however, both steps can be run within the same job, since large tests with many ranks are not required for most development work.

Two primary methods of comparison are used to determine the success of a system test. These are either a direct cell-by-cell comparison of the results for each field using the floating point comparison tools described above (Section \ref{sec:fp-comparing}), or a calculation of the L2 norm of the L1 error vector as described in Section \ref{sec:lwc}. The cell-by-cell comparison is quite accurate, but can be fragile on some complex tests if a small number of cells have errors that are slightly larger than typical, which can lead to false failures when comparing results between systems or compilers. The L2 norm method is less fragile to small errors in a handful of cells, but is generally less sensitive, so we only use it on the tests where it is required, namely the MHD blast wave (\autoref{sec:mhd-blast}) and advecting field loop (\autoref{sec:afl}).

\subsection{Automated Testing}

To ensure that these tests are run regularly and all new code is tested we have made the existing tests as easy to run as possible, and we require that they are all run automatically on each pull request. To facilitate this, Cholla's build directory includes a script which performs all required setup, installs GoogleTest (if requested), builds Cholla, and then runs all the tests. The script also includes a function that combines all of these into a single function call for ease of use, for example, if a user is running tests manually (say, prior to submitting a PR).

Implementing automated testing for HPC codes is not always an easy task. Automated tests are an aspect of Continuous Integration (CI) -- the practice of automating the addition of code changes and additions from a team of developers into a software project. Cholla is currently designed to run on CUDA or HIP capable GPUs, so GPU hardware is required in order to incorporate continuous integration (CI), something that few current CI services offer at a reasonable cost for academic users. Many of Cholla's physics modules are turned on and off at compile time, which presents another challenge, since not all code will be tested for every compilation configurations. Similarly, we need to be able to ensure that Cholla works properly on both AMD and NVIDIA GPUs, which require different compilation targets. This means that there is no single binary file that contains all of Cholla. Instead we have multiple ``builds'' that each require testing. While this does result in a high performance executable that only contains the necessary code, it makes testing the code much more complex due to the number of possible build configurations.

Our solution to these issues was to use a mix of two different systems. When a pull request is submitted to the Cholla GitHub repository several jobs are launched: A GitHub Actions\footnote{\url{https://github.com/features/actions}} job to check code formatting, a GitHub Actions matrix job to build all the HIP/AMD builds using a Docker container, and a Jenkins\footnote{\url{https://www.jenkins.io}} matrix job running on local hardware that runs the CUDA/NVIDIA builds, tests, and static analyzers. Thus, every common configuration of Cholla is built with both CUDA and HIP on every pull request and the CUDA builds are also tested to ensure that no existing or new tests fail.

\section{Summary}
\label{sec:summary}

We have presented the MHD extension to Cholla, a massively parallel, GPU native, astrophysical simulation code. MHD in Cholla uses the Van Leer plus Constrained Transport MHD integrator (VL+CT) \citep{stone_2009}, the HLLD Riemann solver \citep{hlld_2005}, and includes multiple high order reconstruction methods to model numerical solutions to the Eulerian ideal MHD equations on a static mesh.

We showed the modifications required to implement MHD on GPUs compared to CPUs and discussed challenges working within the limits of GPUs in Section \ref{sec:gpu-vs-cpu}. One major challenge was moving computational work and data storage from the CPU to the GPU. Previous versions of Cholla did some of the computation on the CPU and used CPU memory to effectively expand GPU memory. This required regularly copying data between the CPU and the GPU which became a performance bottleneck. The current version of Cholla maintains all the data, and the vast majority of the work, on the GPU which has led to a considerable speedup. As demonstrated in Section \ref{sec:mhd-perf-tests}, these optimizations combined with the highly parallel nature of GPUs make MHD in Cholla extremely fast, with 259 million cell updates per GPU-second when running on a single NVIDIA A100. Cholla also demonstrates excellent weak scaling, and achieves a weak scaling efficiency of 80\% when scaled up to 74,088 GPUs on \textit{Frontier} which utilizes AMD MI250X GPUs; with a total of 14.0 trillion cell updates per second (see Figure \ref{fig:scaling-weak-efficiency}).

We have also presented a suite of canonical MHD tests in Section \ref{sec:mhd-tests}. These tests demonstrate the accuracy of Cholla across a broad range of problems. They also demonstrate that the VL+CT MHD algorithm does an excellent job of maintaining the divergence free condition even in highly challenging settings.

To accommodate the increasing complexity of Cholla and facilitate multiple simultaneous development efforts, we have also added a structured testing framework, described in Section \ref{sec:testing}. This framework is based on GoogleTest, augmented with custom testing tools. This approach enables running tests that range from single function unit tests to massively parallel system tests across the scale of an entire cluster within the same framework. We have further integrated these tests with GitHub Actions to run formatting, static analysis, and builds with various physics configurations along with Jenkins running on local resources.

Cholla is free and open source software available at \url{https://github.com/cholla-hydro/cholla}. In general, Cholla is designed to be flexible and modular, and can be run with or without the new MHD module. In addition, the Cholla framework can easily accommodate additional physics modules in the future, some of which are in progress. We welcome new development efforts, and hope that this work will be a resource for the broader astrophysics simulation community.

\section{Acknowledgements}

RVC thanks Alwin Mao, Helena Richie, Orlando Warren, Kyle Felker, and Seth Cook for many helpful discussions. Special thanks to Daniel Perrefort at the Center for Research Computing for his help in setting up and maintaining the Jenkins environment for our automated testing. This research was supported in part by the University of Pittsburgh Center for Research Computing, RRID:SCR\_022735, through the resources provided. Specifically, this work used the H2P cluster, which is supported by NSF award number OAC-2117681. This research also used resources of the Oak Ridge Leadership Computing Facility, which is a DOE Office of Science User Facility supported under Contract DE-AC05-00OR22725, using Frontier CAAR allocations CSC380 and AST181. E.E.S. acknowledges support from NASA TCAN grant 80NSSC21K0271, NASA ATP grant 80NSSC22K0720, StScI grant HST-AR-16633.001-A, and the David and Lucile Packard Foundation (grant no. 2022-74680).

\software{
Cholla \citep{schneider_2015, caddy_2024},
numpy \citep{van2011numpy},
matplotlib \citep{hunter2007matplotlib},
HDF5 (\href{https://github.com/HDFGroup/hdf5}{The HDF Group. Hierarchical Data Format, version 5.}),
GoogleTest (\url{https://github.com/google/googletest})
}

\bibliography{bibliography.bib}

\end{document}